\documentclass[preprint,12pt]{elsarticle}

\usepackage{amssymb}
\usepackage{amsmath}
\usepackage{enumitem}
\usepackage{xurl}
\usepackage{multirow}
\usepackage{tabularx}
\usepackage{graphicx}

\journal{Wind Engineering and Industrial Aerodynamics}

\begin{document}

\begin{frontmatter}



\title{Hurricane Laura-induced loads on wind turbines -- Part I: High-fidelity modeling and sensitivity to turbulent inflow conditions}

\author[label1,label2]{Miguel Sanchez-Gomez} 
\author[label1,label3]{Lucas Carmo} 
\author[label1]{Matthew Churchfield} 
\author[label1]{Jason Jonkman} 
\author[label1]{Walt Musial}

\affiliation[label1]{organization={National Laboratory of the Rockies},
            addressline={15013 Denver West Parkway}, 
            city={Golden},
            postcode={80401}, 
            state={Colorado},
            country={United States}}
\affiliation[label2]{organization={Johns Hopkins University},
            addressline={3400 North Charles Street}, 
            city={Baltimore},
            postcode={21218}, 
            state={Maryland},
            country={United States}} 
\affiliation[label3]{organization={University of S\~ao Paulo},
    addressline={Cidade Universit\'aria},
            city={S\~ao Paulo},
    postcode={05508-000},
            state={SP},
            country={Brazil}}

\begin{abstract}
Tropical cyclones can produce extreme winds in the lowest portion of the boundary layer, posing high risks to critical energy infrastructure, including wind energy systems. This study evaluates the dynamic response of an idling wind turbine to turbulent inflow generated during the passage of a hurricane. We performed large-eddy simulations of Hurricane Laura (2020) at several grid resolutions -- coarse $(\Delta x=55.55\,{\rm m})$, intermediate $(\Delta x=25\,{\rm m})$, and fine $(\Delta x=10\,{\rm m})$ -- to investigate differences in mean flow characteristics and turbulence structures as the hurricane progresses across a virtual wind turbine location. The large-eddy simulation data served as direct inflow conditions for aeroelastic simulations to assess the turbine's response. Our results reveal significant differences in the mean wind and turbulent flow structures between the grid resolutions, leading to distinct impacts on the wind turbine's aeroelastic behavior. These findings emphasize the need for high-resolution inflow information to better predict extreme loads and inform the design of wind turbines in regions frequently affected by tropical cyclones. As a continuation of this work, Part II of this study will evaluate the turbine's response from different locations relative to the storm's trajectory and compare it with current wind turbine design practices.
\end{abstract}

\begin{graphicalabstract}
\end{graphicalabstract}

\begin{highlights}
\item High-fidelity turbulence-resolving simulations of a historical hurricane are used to assess wind turbine loads during storm passage.
\item Near-surface mean wind and turbulence characteristics are sensitive to grid spacing, influencing the resulting wind turbine response.
\item Sensitivity to turbulent inflow conditions highlights the grid-spacing requirements for converged loads on idling wind turbines.
\end{highlights}


\end{frontmatter}

\section{Introduction}  
Tropical cyclones represent a significant hazard to critical energy infrastructure, including wind energy systems. Wind energy infrastructure is now present in a wider range of environmental and climatic settings, including areas affected by tropical cyclones. Furthermore, the recurrence period of tropical cyclones in some locations can be smaller than the wind farm lifetime (e.g., 25 years) \citep{hallowell_hurricane_2018,keim_spatiotemporal_2007,pandey_typhoon_2022,chen_assessing_2018,huang_long-term_2019,fang_estimation_2020}, calling for a more thorough investigation of the hazard associated with installing and operating wind turbines in tropical cyclone-prone areas. Several high-intensity storms have already impacted wind turbines in the North Atlantic and Pacific regions, causing catastrophic damage to the blades, towers, and foundations. In 2003, Typhoon Maemi moved over Miyakojima Island in Japan, impacting six wind turbines with estimated mean hub-height wind speeds of $60\,{\rm m\,s^{-1}}$ \citep{ishihara2005analysis,takahara2004damages}. Three of the six wind turbines collapsed, and the other three suffered destructive damage to the blades \citep{ishihara2005analysis,takahara2004damages}. In 2017, Hurricane Maria made landfall in Puerto Rico, causing widespread damage to the island's energy infrastructure \citep{kwasinski_effects_2018,kwasinski_hurricane_2019}. The northeast quadrant of Hurricane Maria impacted the Punta Lima wind farm, causing catastrophic damage to the towers and blades of 13 wind turbines. Turbines in a wind farm farther away from the hurricane's path were undamaged, highlighting the complex nature of hurricane wind fields and their effect on wind energy systems. Additional incidents in China (typhoons Ragasa and Yagi) also caused the destruction of wind turbines as they made landfall. 

Damage to infrastructure may result from design specifications that do not fully represent the extreme winds in tropical cyclones. Engineering standards for energy and civil infrastructure employ simplified wind models that do not capture the complex flow features in tropical cyclones to represent extreme-event wind loading during structural design. The American Society of Civil Engineers (ASCE) provides guidance for the design of structures that are subject to building code requirements \citep{asce_minimum_2021}. Similarly, for offshore energy infrastructure, the American Petroleum Institute (API) publishes recommended practices for the design of offshore structures used in oil and natural gas drilling and production \citep{api_recommended_2000,api_interim_2007}. The International Electrotechnical Commission (IEC) also provides guidelines for the design of land-based and offshore wind turbines \citep{iec-1_wind_2019,iec-3_wind_2019}. These standards propose different formulations for the extreme wind speed inflow models; API proposes a logarithmic model, while the ASCE and IEC propose a power-law model. However, their approach to determining the shape and magnitude of the extreme wind profile follows the same  assumptions of near-neutral atmospheric stability and sole dependence on the surface shear stress. Furthermore, turbulence in these models may not be representative of tropical cyclone conditions. For instance, the Kaimal turbulence model \citep{kaimal_1972}, developed using extensive observations in flat terrain, assumes unidirectional and horizontal mean winds that only vary vertically, with stationary and horizontally homogeneous turbulence. However, winds in the lowest $300\,{\rm m}$ of the tropical cyclone boundary layer are characterized by large shear (vertical and horizontal), and turbulent eddies are also highly anisotropic \citep{sroka_organized_2021,oguejiofor_role_2024,sanchezgomez_turbulenceresolving_2025}. The Mann uniform shear model \citep{mann_spatial_1994} incorporates uniform vertical shear of the streamwise wind, distorting the isotropic energy spectrum. However, wind fields in tropical cyclones are sheared horizontally and vertically, and the mean vertical and cross-stream components of the wind vector are non-zero, likely modifying turbulence. In addition, the Mann model may not capture the energy at low frequencies in intense storms \citep{muller_can_2025}.

Quantifying the potential impact of tropical cyclones on wind energy systems requires accurate wind fields representative of extreme storms. The extreme wind models used in wind turbine design do not fully consider the mean wind structure and turbulence in tropical cyclones \citep{worsnop_gusts_2017,sanGom_2023}. Deficiencies in the inflow models used for wind turbine design can yield inaccurate estimates on the associated loads during tropical cyclones \citep{kapoor_hurricane_2020}. Improving the engineering wind models used for structural design to capture flow features in tropical cyclones requires accurate characterization of mean wind and turbulent conditions in historical storms.

Characterizing winds at turbine-relevant heights during extreme storms is difficult due to the complex nature of tropical cyclones. Tropical cyclones have large horizontal and vertical gradients \citep{willoughby_concentric_1982,marks_structure_2008,lorsolo_estimation_2010,zhang_asymmetric_2013} that modify turbulence near the surface \citep{ren_basic_2022,sanchez_gomez_toward_2025,sanchezgomez_turbulenceresolving_2025} and at upper levels \citep{lorsolo_estimation_2010}. Moreover, different storms can have different vertical wind structures \citep{franklin_gps_2003,zhang_characteristic_2011,wang_long-term_2015}, adding variability to the range of turbulent and mean wind conditions that turbines can experience near the surface. The extreme conditions in tropical cyclones, combined with observational undersampling \citep{nolan_limits_2014} and their large spatial extent, make them difficult to characterize. As a result, wind measurements needed to describe turbulence at turbine-relevant heights and near the eyewall remain extremely limited.

High-fidelity numerical simulations of extreme storms can provide critical information on turbulence during tropical cyclones and inform wind turbine design. High-fidelity turbulence-resolving numerical simulations of tropical cyclones can be used to procure the data for the mean wind speed profiles and turbulence characteristics of a hurricane boundary layer. Large-eddy simulations (LES) of idealized \citep{bryan_simple_2017,worsnop_gusts_2017,ren_largeeddy_2020,sanGom_2023} and historical \citep{li_effects_2021,sanchezgomez_turbulenceresolving_2025} storms have been used to study the complex wind fields in tropical cyclones. In addition, wind field statistics (i.e., mean wind profiles and turbulence intensity) derived from idealized simulations can be used alongside synthetic turbulence models to estimate turbine loads during extreme events \citep{kapoor_hurricane_2020}. Although synthetic turbulence models are commonly used in wind engineering, they may not represent the diversity of observable spectra and may need fine tuning \citep{keith_learning_2021}.






Here, we assess the sensitivity of a wind turbine's response to realistic turbulent inflow across different spatiotemporal resolutions during the passage of a hurricane. To this end, we conduct microscale simulations of a high-intensity storm at varying grid spacing to examine differences in the turbulent wind fields. Then, we use the high-fidelity LES data as direct inflow to an aeroelastic wind turbine simulation to examine the wind turbine's response to the wind inflow obtained from the different grid resolutions. Part II of this study examines the importance of simulation length, turbine location relative to the storm track, and comparison to wind turbine design standards. The remainder of the article is structured as follows. Section \ref{sec:cfd_framework} describes the numerical framework used for generating the hurricane wind fields, and the aeroelastic wind turbine simulations are described in Sect. \ref{sec:aeroelast_sims}. We study the hurricane wind fields in Sect. \ref{sec:hurr_winds} and the wind turbine's response in Sect. \ref{sec:windturb_response}. Lastly, conclusions and future work are presented in Sect. \ref{sec:concl}. 

\section{Numerical modeling framework of a historical hurricane \label{sec:cfd_framework}}
We conduct mesoscale-microscale coupled simulations of Hurricane Laura to study the wind turbine's response to a hurricane passage. The large-scale structure of hurricane Laura develops in a precursor two-domain mesoscale simulation using the Weather Research and Forecasting (WRF) model \citep{skamarock_description_2019}, as explained in Sanchez-Gomez et al. \cite{sanchezgomez_turbulenceresolving_2025}. The outer mesoscale domain uses a horizontal grid spacing $\Delta x$ of $4.5\,{\rm km}$ with spectral nudging, while a vortex-following domain with $\Delta x=1.5\,{\rm km}$ is used to improve the storm's representation as it moves across the North Atlantic (Table \ref{tab:numConfig}). Atmospheric conditions from the National Centers for Environmental Prediction Global Forecast System (GFS) analysis \citep{cisl_rda_dsd084001} and sea surface temperature from the Operational Sea Surface Temperature and Ice Analysis \citep{stark_ostia_2007} are used for initial and boundary forcing of the outer mesoscale domain. This two-domain framework accurately captures the intensity, track, and vertical structure of Hurricane Laura \citep{sanchezgomez_turbulenceresolving_2025}. The mesoscale simulation employs a stretched vertical grid, where $\Delta z$ increases linearly from $\Delta z=8.5\,{\rm m}$ to $\Delta z=13\,{\rm m}$ between the surface and $z=150\,{\rm m}$; then, $\Delta z$ increases linearly to  $\Delta z=33\,{\rm m}$ at $z=300\,{\rm m}$. Farther aloft, $\Delta z$ increases by approximately $6\,{\rm m}$ for every $100\,{\rm m}$ gained in elevation.

\begin{figure}[htbp!] 
\centering \includegraphics[height=7cm]{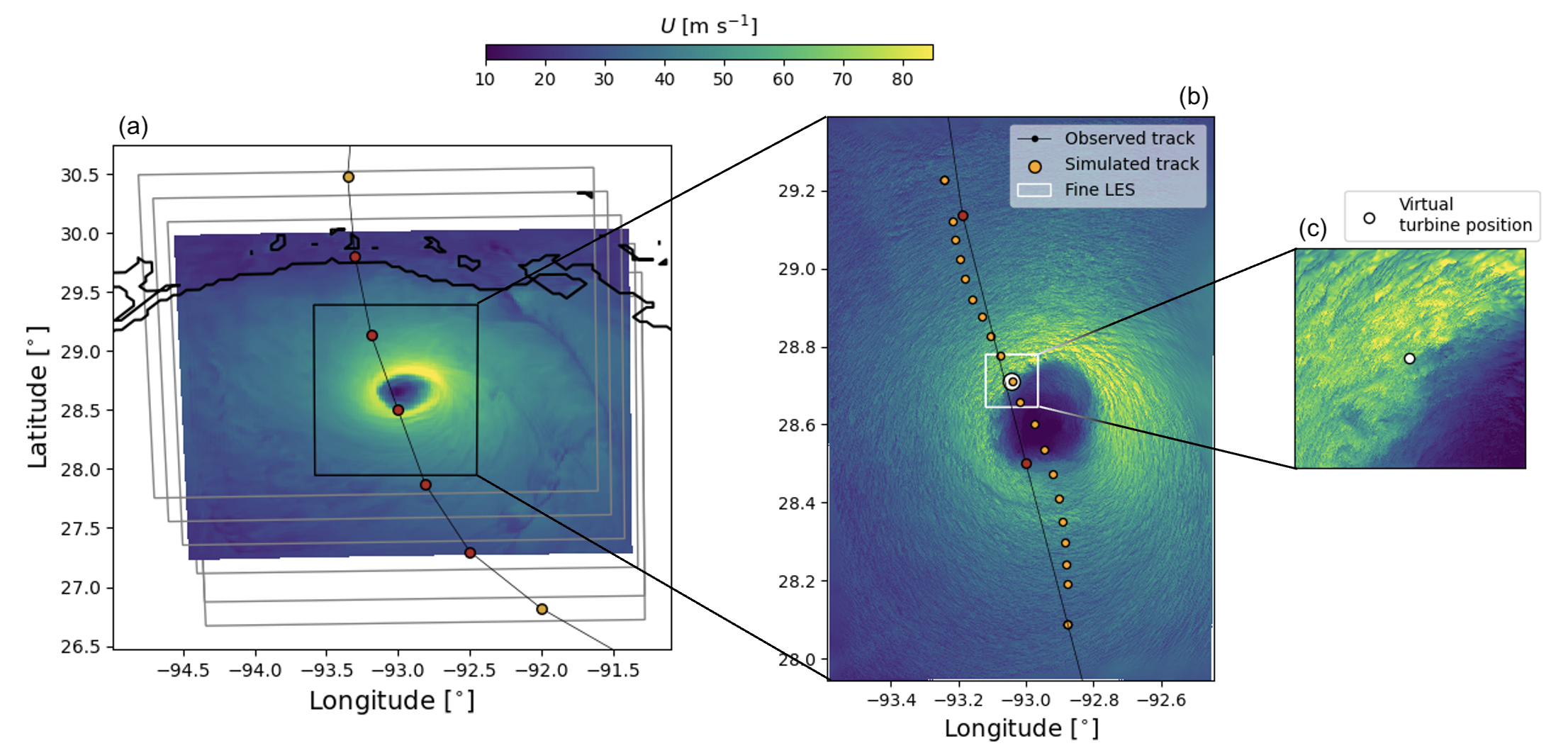}
\caption{Numerical domain configuration for the mesoscale-microscale coupling of Hurricane Laura. Data from the precursor vortex-following mesoscale domain (a) provide the large-scale structure of the storm and its temporal evolution to a coarse LES (b). An additional refinement region (c) captures turbine-scale turbulence as the hurricane moves northward.}
\label{fig:meso_to_micro}
\end{figure}

Microscale simulations of Laura are conducted using an offline coupling framework, where high-frequency mesoscale output from the vortex-following domain is used to inform the turbulence-resolving simulations. Three-dimensional (3D) output every 30 s from the vortex-following domain between 26 August 2020 22:00 UTC and 27 August 2020 04:00 UTC is used as initial and boundary conditions for the microscale simulations using the Moving-Mesoscale to Static-LES Integrated Coupling (MOSAIC) framework \citep{sanchezgomez_turbulenceresolving_2025}. Figure \ref{fig:meso_to_micro} illustrates the mesoscale-to-microscale coupling employed here, where the 3D output from the vortex-following grid (Fig. \ref{fig:meso_to_micro}a) provides initial and boundary conditions to a fixed coarse-resolution LES (Fig. \ref{fig:meso_to_micro}b), which in turn informs a higher-resolution LES (Fig. \ref{fig:meso_to_micro}c). As shown in Sanchez-Gomez et al. \cite{sanchezgomez_turbulenceresolving_2025}, this coupling framework ensures that the storm's horizontal and vertical structure, intensity, and track are well captured by microscale simulations. The numerical configuration of the mesoscale and microscale domains employed here is shown in Table \ref{tab:numConfig}.

A coarse-resolution LES of Laura captures large-scale turbulence as the hurricane moves northward (Fig. \ref{fig:meso_to_micro}b). The coarse-resolution LES encompasses the entire eyewall of the storm and the flow field far from the storm's center, enabling turbulence analyses at different radial and azimuthal locations relative to the storm's center and to its trajectory. Initial and boundary conditions from the precursor mesoscale simulation ensure that the storm's evolution and intensity are well represented in the coarse LES, while turbulence develops within the domain by using a horizontal grid spacing $\Delta x=55.55\,{\rm m}$. The coarse LES domain uses the same vertical grid as the mesoscale, where $\Delta z<\Delta x$ in the lowest $700\,{\rm m}$ of the boundary layer enables turbulence structures to develop near the surface. The coarse-resolution LES is run using the WRF model \citep{skamarock_description_2019}, further ensuring that the large-scale flow remains dynamically consistent with the mesoscale boundary conditions.

\begin{table*}[t]
\caption{Numerical domain configuration of the mesoscale and microscale simulations, including the number of grid points along each $i$-direction $(n_i)$, domain size $(L_i)$, horizontal grid spacing $(\Delta x)$, mean vertical grid spacing across the turbine rotor layer $(\Delta z_{\rm RL})$, and time step $(\Delta t)$. } %
\resizebox{\textwidth}{!}{
\begin{tabular}{lccccccc}
\hline
\hline
Simulation & Numerical Model & Domain & $(n_x,n_y,n_z)$ $[-]$ & $(L_x,L_y,L_z)$ $[\rm km]$ & $\Delta x$ $[\rm m]$ & $\langle \Delta z \rangle_{\rm RL}$ $[\rm m]$  & $\Delta t$ $[\rm s]$ \\
\hline
\multirow{2}{*}{Mesoscale} & \multirow{2}{*}{WRF} & M01 & $(400,400,79)$ & $(1800,1800,16)$ & $4500$ & \multirow{2}{*}{$14$}  & $15$ \\ %
 &  & M02 & $(205,205,79)$ & $(307.5,307.5,16)$ & $1500$ &    & $3.75$ \\ %
\hline
\multirow{3}{*}{Microscale} & WRF & Coarse & $(2000,2890,79)$ & $(111.1,160.5,16)$ & $55.55$ & $14$  & $0.01$  \\ %
 & FastEddy & Intermediate & $(600,602,58)$ & $(15,15.05,1)$ & $25$ & $10$  & $0.025$ \\ %
 & FastEddy & Fine & $(1500,1498,90)$ & $(15,14.98,1)$ & $10$ & $5$  & $0.01$ \\ %
\hline
\hline
\end{tabular}
}
\label{tab:numConfig}
\end{table*}

We conduct turbine-scale LES of Hurricane Laura to evaluate the sensitivity of the wind turbine's response to the turbulent wind field. The refinement region is located at the storm's center halfway through the duration of the simulation (Fig. \ref{fig:meso_to_micro}c). Turbine-scale simulations at two different grid spacings are conducted using the GPU-resident FastEddy\textsuperscript{\textregistered} model \citep{sauer_fasteddy_2020,munozesparza_fasteddy_2022}, which enables efficient, high-resolution, and long-duration LES. An intermediate-resolution simulation with horizontal grid spacing $\Delta x=25\,{\rm m}$ and a fine-resolution simulation with $\Delta x=10\,{\rm m}$ are run to study sensitivity to grid resolution. The vertical grid spacing in the intermediate-resolution simulation is $\Delta z=7.5\,{\rm m}$ at the surface stretching to $\Delta z=35\,{\rm m}$ at the domain top. For the fine-resolution LES, the vertical grid spacing is $\Delta z=5\,{\rm m}$ at the surface stretching to $\Delta z=25\,{\rm m}$ at the domain top. Additional simulations with finer grid spacing $(\Delta x=5\,{\rm m})$ confirmed that turbulent conditions and the turbine's response to the hurricane are adequately represented with the $\Delta x=10\,{\rm m}$ LES (see Appendix \ref{sec:appA}). Three-dimensional output from the coarse LES at $1\,{\rm Hz}$ is used as initial and boundary conditions for the turbine-scale simulations between 26 August 2020 22:30 UTC and 27 August 2020 04:00 UTC. Note that the vertical domain size of the turbine-scale LES is $1\,{\rm km}$, but the hurricane's boundary layer may extend farther aloft; therefore, we prescribe boundary conditions at the top of the domain from the coarse LES so that the storm's structure remains well represented throughout the entire boundary layer.

Physical processes in the atmosphere are parameterized in a similar manner in the coarse LES and mesoscale simulations, except for turbulent mixing. Both simulations model cloud microphysics using the single-moment 6-class cloud physics scheme \citep{hong_wrf_2006}, longwave radiation effects using the rapid radiative transfer model \citep{mlawer_radiative_1997}, and shortwave radiation effects using the Dudhia model \citep{dudhia_numerical_1989}. Because we are using WRF for the coarse-resolution LES and FastEddy for the turbine-scale LES, some physical parameterizations may differ between the microscale simulations. FastEddy uses a simple scheme for representing microphysical processes, only considering the transfer between water vapor and liquid water (cloud) species \citep{munozesparza_fasteddy_2022}. Furthermore, the turbine-scale simulations do not include radiation. Both WRF and FastEddy employ Monin--Obukhov similarity theory to estimate surface fluxes \citep{jimenez_revised_2012}. The surface-exchange coefficient for momentum $(C_D)$ is capped at $C_D\approx2.4\times10^{-4}$ following Donelan \cite{donelan_limiting_2004}, while the surface fluxes for heat and moisture are parameterized following Garratt \cite{garratt_atmospheric_1999}. Although some physical processes may be parameterized differently in FastEddy compared to WRF, the small domain size of the turbine-scale simulations and high-frequency boundary forcing from the coarse LES ensure that the structure of the storm is consistent between all simulations.

High-frequency output at the center of the microscale domains are used to evaluate wind conditions during the hurricane passage. The 3D velocity fields are saved at $1\,{\rm Hz}$, $2\,{\rm Hz}$, and $4\,{\rm Hz}$ in the coarse-, intermediate-, and fine-resolution LESs, respectively. A higher-frequency output is used for the $\Delta x=25\,{\rm m}$ and $\Delta x=10\,{\rm m}$ simulations, as the finer grids resolve smaller eddies in the flow than the coarse grids do. We examine the hurricane evolution at a virtual turbine position located at the center of the LES domains, which also matches the location of the eye of the storm at the midpoint of the microscale simulations (Fig. \ref{fig:meso_to_micro}b,c). For the subsequent analysis, we consider the International Energy Agency Wind Technology Collaboration Programme (IEA Wind) 15 MW reference wind turbine \citep{gaertner2020iea} with a hub height $z_h=150\,{\rm m}$ and a rotor diameter $D=240\,{\rm m}$.

\subsection{Turbulence development}
We use the cell-perturbation method \citep{munoz-esparza_bridging_2014,munoz-esparza_stochastic_2015} to reduce the fetch and time required to develop turbulence in the microscale simulations. Random $\pm 0.5\,{\rm K}$ are added in the lowest $1\,{\rm km}$ of the boundary layer near the lateral boundaries of the coarse LES domain to instigate 3D turbulence development. We also add random temperature perturbations to the lateral inflow boundaries of the higher-resolution LES domains to reduce the fetch required to resolve the smaller scales of turbulence within the finer grid. To evaluate the development of turbulence across the microscale domains, we consider the temporal/spatial evolution of the energy spectra at $z=150\,{\rm m}$ in the vicinity of the virtual turbine, where the fluctuation component of the velocity field is estimated via Reynolds decomposition using a 10 min moving average. 

\begin{figure}[htbp!] 
\centering \includegraphics[height=5cm]{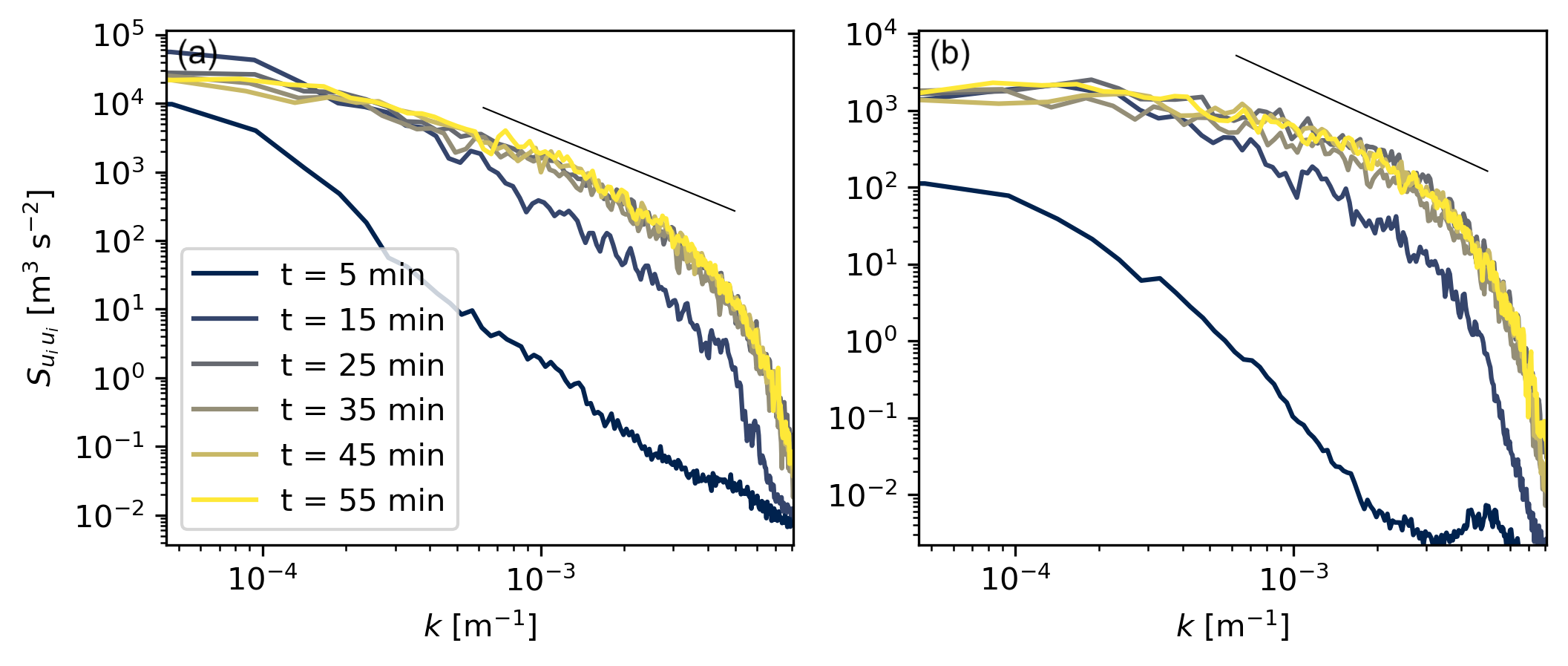}
\caption{Temporal evolution of energy spectra of the horizontal wind speed magnitude (a) and vertical wind component (b) at $z=150\,{\rm m}$ for the coarse-resolution $(\Delta x=55.55\,{\rm m})$ LES of Hurricane Laura. Turbulence spectra are estimated using the time-series output at the center of the LES domain and converted from frequency to wavenumber space using Taylor's frozen turbulence hypothesis.}
\label{fig:turbSpinup_coarseLES}
\end{figure}

Turbulence at the center of the coarse LES domain develops shortly after initialization (Fig. \ref{fig:turbSpinup_coarseLES}). The one-dimensional energy spectra of the horizontal (Fig. \ref{fig:turbSpinup_coarseLES}a) and vertical (Fig. \ref{fig:turbSpinup_coarseLES}b) wind speed converge 25 minutes after initialization, indicating turbulence has propagated across all resolvable scales. The energy spectra show very low energy content in the flow at high wavenumbers, suggesting the coarse grid is not capable of adequately resolving the small scales of turbulence associated with wavenumbers $k\gtrapprox2\times10^{-3}\,{\rm m^{-1}}$. Using Taylor’s frozen-turbulence hypothesis, which relates temporal and spatial scales via $f\approx\overline{U}k$, the coarse LES cannot resolve eddies with characteristic frequencies $f\gtrapprox0.1\,{\rm Hz}$. Because the flow at hub height becomes fully turbulent 25 minutes after initialization, the first 30 minutes of simulation time are discarded for turbulence spin-up.

\begin{figure}[htbp!] 
\centering \includegraphics[height=8.5cm]{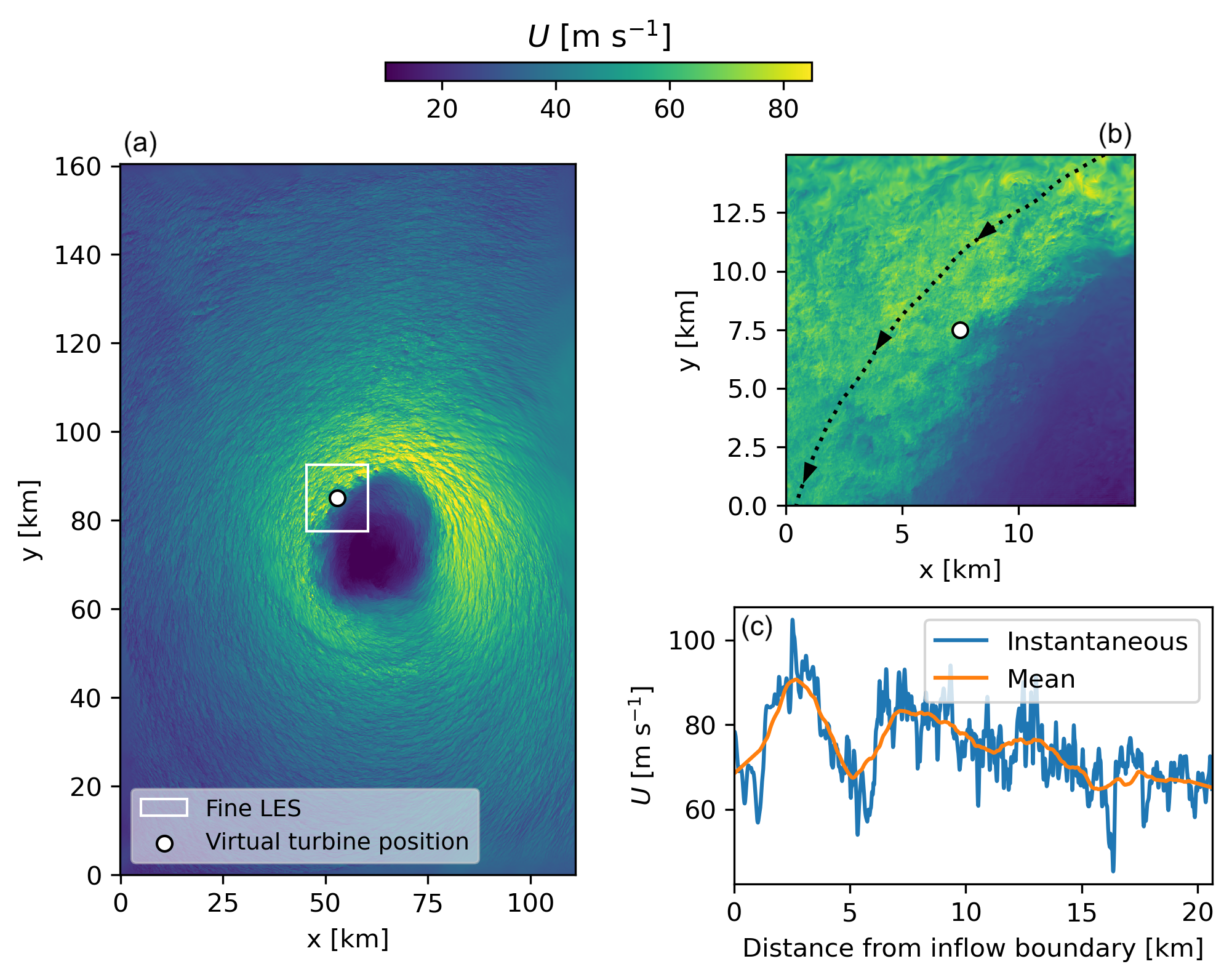}
\caption{Plan view of the horizontal wind speed field in the coarse-resolution LES (a) and the fine-resolution LES (b). Panel (c) shows the wind speed magnitude along the streamline shown in panel (b) as a function of distance from the lateral inflow domain boundary. }
\label{fig:sl_explain}
\end{figure}

The turbine-scale LES domains are initialized when turbulence is fully developed in the coarse LES domain. Initial and boundary conditions from the coarse LES already provide fully developed large-scale turbulence; however, because of the grid refinement between the coarse and higher-resolution simulations, smaller-scale eddies are expected to form within the finer grids. We evaluate the development of smaller scales of turbulence using the spatial evolution of the energy spectra of the horizontal wind speed for the $\Delta x = 10\,{\rm m}$ simulation. The horizontal wind speed is sampled along streamlines emanating from the domain inflow boundaries (Fig. \ref{fig:sl_explain}b). We obtain the fluctuation velocity along each streamline using Reynolds decomposition, where the mean velocity is estimated using a moving average along the streamline with an averaging window of $2\,{\rm km}$. Each streamline emanating from the inflow boundaries is divided into $2\,{\rm km}$ long segments, and we estimate the spectral density of each segment. A low-pass Butterworth filter is applied to the sampled velocity field along each streamline with cutoff wavenumber at $k_c=1/2\Delta x$ to prevent aliasing.

\begin{figure}[htbp!] 
\centering \includegraphics[height=5cm]{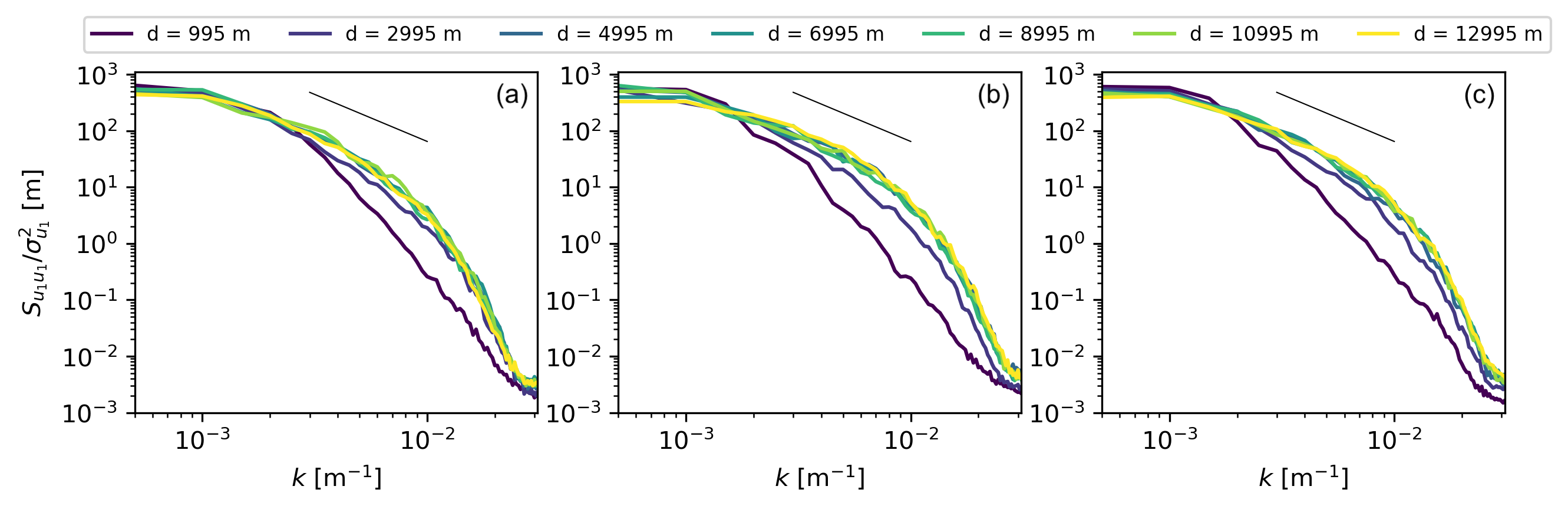}
\caption{Energy spectra of the horizontal wind speed when the eye of the storm is far south of the refinement region (a), when the eyewall is within the refinement region (b), and when the eye of the storm is far north of the refinement region (c).}
\label{fig:turbSpinup_fineLES}
\end{figure}

Small-scale turbulence in the turbine-scale LESs develops within the first $4$--$5\,{\rm km}$ from the domain inflow boundaries (Fig. \ref{fig:turbSpinup_fineLES}). The inflow boundaries to the high-resolution LES domain change as the hurricane moves northward. At initialization, when the hurricane's eye is far south of the refinement region, the inflow to the high-resolution LES domain is predominantly from the eastern boundary, and turbulence fully develops after $3\,{\rm km}$ of fetch (Fig. \ref{fig:turbSpinup_fineLES}a). At the midpoint of the simulation, when the storm's eyewall reaches the refinement region, winds from the inflow boundary are fastest. As a result, small-scale turbulence requires about $5\,{\rm km}$ of fetch to develop (Fig. \ref{fig:turbSpinup_fineLES}b). In the later stages of the simulation, when the eye of the hurricane is far north of the refinement region, the inflow to the high-resolution LES is primarily from the north and west boundaries, and turbulence again requires small fetch (Fig. \ref{fig:turbSpinup_fineLES}c). Because the inflow boundaries change over time, the high-resolution LES domain must be large enough for turbulence to fully develop inward from any of the lateral domain boundaries. In a conservative manner, our horizontal domain size exceeds $L_x,L_y>10\,{\rm km}$ in size to ensure that small-scale eddies are always fully developed at the domain's center.
 
\section{Aeroelastic simulation of wind turbine response \label{sec:aeroelast_sims}}
We conduct aeroelastic simulations of the IEA Wind 15 MW reference turbine \citep{gaertner2020iea} atop a monopile under idling conditions to evaluate the turbine's response to the storm's passage. The aeroelastic simulations are conducted using OpenFAST, an engineering software for coupled nonlinear aero-hydro-servo-elastic simulations of wind turbines in the time domain. The OpenFAST inputs to the reference turbine are publicly available and summarized in Table \ref{tab:turbine_properties}. Our simulations focus on sensitivity to wind conditions, as such, we do not include hydrodynamic loading on the substructure. In this way, the OpenFAST modules active in our simulations are ElastoDyn, BeamDyn, AeroDyn, ServoDyn, and InflowWind.

\begin{table}[t]
    \centering
    \caption{General characteristics of the IEA Wind 15 MW reference wind turbine.}\label{tab:turbine_properties} 
    \begin{tabular}{lr}
        \hline
        \hline
        Rated power & $15\,{\rm MW}$ \\
        Hub height & $150\,{\rm m}$ \\
        Rotor diameter $(D)$ & $240\,{\rm m}$ \\
        Hub diameter & $7.94\,{\rm m}$ \\
        Cut-in wind speed & $3\,{\rm m\,s^{-1}}$\\
        Cut-out wind speed & $25\,{\rm m\,s^{-1}}$\\
        Transition piece height & $15\,{\rm m}$ \\
        First tower fore-aft mode & $0.17\,{\rm Hz}$ \\
        First blade flapwise mode & $0.56\,{\rm Hz}$ \\
        \hline
        \hline
    \end{tabular}%
\end{table}%

Wind turbine blades are modeled using BeamDyn, which allows the modeling of large deflections by using geometrically exact beam theory. The tower, nacelle, and generator are modeled using ElastoDyn, where we consider the first and second fore–aft and side-to-side tower bending modes. Because hydrodynamic loads on the monopile are neglected, we do not model the monopile using SubDyn. Instead, the ElastoDyn input file is modified so that the natural frequencies and mode shapes of the clamped tower match those of the tower supported by the monopile. To represent turbine conditions during a high-intensity storm, the turbine blades are feathered by setting their pitch angle to $90^\circ$. To model the idling turbine in ServoDyn, blade pitch control is disabled and the generator startup time exceeds that of the simulation, effectively disabling it. Here, we assume a battery backup system is available during the hurricane passage and thus the yaw control is active to align the rotor with the time-varying wind direction using a yaw rate of $0.25^\circ\,{\rm s^{-1}}$ and an error threshold of $8^\circ$. To compute aerodynamic forces on the structure, we use AeroDyn using a lookup table (without unsteady airfoil aerodynamics) with airfoil polars calibrated to computational fluid dynamics results in deep stall.

Turbulent inflow from the microscale simulations is used to evaluate the wind turbine's response to the hurricane's passage using InflowWind. The velocity fields are sampled over a volume of size $(L_x,L_y,L_z)=(300\,{\rm m},300\,{\rm m},300\,{\rm m})$, fully encompassing the IEA Wind 15 MW turbine. To ensure turbulence is fully developed, the sampling volume is located at the center of the turbine-scale LES domains. The three-dimensional velocity fields at each grid point within the volume are saved at $1\,{\rm Hz}$, $2\,{\rm Hz}$, and $4\,{\rm Hz}$ for the $\Delta x=55.55\,{\rm m}$, $\Delta x=25\,{\rm m}$, and $\Delta x=10\,{\rm m}$ LESs, respectively. The instantaneous velocity fields are interpolated in time to match the time step $(\Delta t = 0.01\,{\rm s})$ of the aeroelastic simulations. 

\section{Wind conditions during the hurricane passage \label{sec:hurr_winds}}
Before investigating the turbine's response to the hurricane passage, we examine differences in the velocity fields between the microscale simulations. We study the near-surface wind fields in a turbine-aligned coordinate system centered on the virtual turbine position. The turbine-aligned coordinate system is composed of the streamwise $(x^*)$, cross-stream $(y^*)$, and vertical $(z)$ axes, with the streamwise direction defined at hub height. The streamwise $(u_1)$ and cross-stream $(u_2)$ wind components are defined using Eqs. \eqref{eq:u1} and \eqref{eq:u2}, where $\overline{\varphi}_h$ is the time-averaged hub-height wind direction at the virtual turbine position, $u$ is the velocity component along the $x$-direction, and $v$ is the velocity component along the $y$-direction. The vertical wind component is the same in the turbine-aligned and Cartesian coordinate system (Eq. \eqref{eq:u3}). We use Reynolds decomposition to study the mean flow and turbulent conditions in the storm, where the mean velocity field $(\overline{u}_i)$ is calculated using a 10 min moving average and the fluctuation velocity components are obtained as $u'_i=u_i-\overline{u}_i$. To examine the temporal evolution of the mean wind and turbulence during the hurricane passage, we focus on four representative times that capture key stages of the storm’s wind-field evolution. First, we analyze winds when the hurricane is far south from the virtual turbine ($t\approx1\,{\rm h}$; purple line in Fig. \ref{fig:meanWindProfiles}c); second, when the leading edge of the eyewall reaches the turbine position ($t\approx2\,{\rm h}$; light-blue line in Fig. \ref{fig:meanWindProfiles}c); third, when the trailing edge of the storm's eyewall impacts the turbine ($t=3.8\,{\rm h}$; pink line in Fig. \ref{fig:meanWindProfiles}c); and fourth, when the hurricane is far north ($t=4.8\,{\rm h}$; yellow line in Fig. \ref{fig:meanWindProfiles}c). For each of the subsequent sections, we examine the wind field structure in the fine-resolution LES and then compare it with the results from the coarse- and intermediate-resolution simulations.

\begin{eqnarray}
u_1(x^*,y^*,z) &=& u(x,y,z)\cos (\overline{\varphi}_h ) + v(x,y,z)\sin (\overline{\varphi}_h ) \label{eq:u1}\\ 
u_2(x^*,y^*,z) &=& -u(x,y,z)\sin (\overline{\varphi}_h ) + v(x,y,z)\cos (\overline{\varphi}_h )
\label{eq:u2} \\ 
u_3(x^*,y^*,z) &=& w(x,y,z)
\label{eq:u3}
\end{eqnarray}

\subsection{Mean wind profiles}
We begin our analysis by examining the mean wind structure near the surface. To study the mean wind profile, we consider the vertical shear of the streamwise wind speed and wind direction across the turbine rotor layer. We quantify vertical wind speed shear using the least-squares fit to the power-law wind profile $U(z)=U_{\rm ref}(z/z_{\rm ref})^\alpha$, where $U_{\rm ref}$ is the wind speed at a reference height $z_{\rm ref}$, here defined as the turbine's hub height $(z_{\rm ref}=150\,{\rm m})$. We quantify vertical wind direction shear as the shortest rotational path between wind vectors at the bottom $z=30\,{\rm m}$ and top $z=270\,{\rm m}$ of the turbine rotor layer. Positive wind direction shear, also known as wind veer, is common in the northern hemisphere and is associated with warm-air advection; whereas negative wind direction shear, also known as wind backing, is less common and is mostly associated with cold-air advection behind frontal passages. 

\begin{figure}[htbp!] 
\centering \includegraphics[height=8cm]{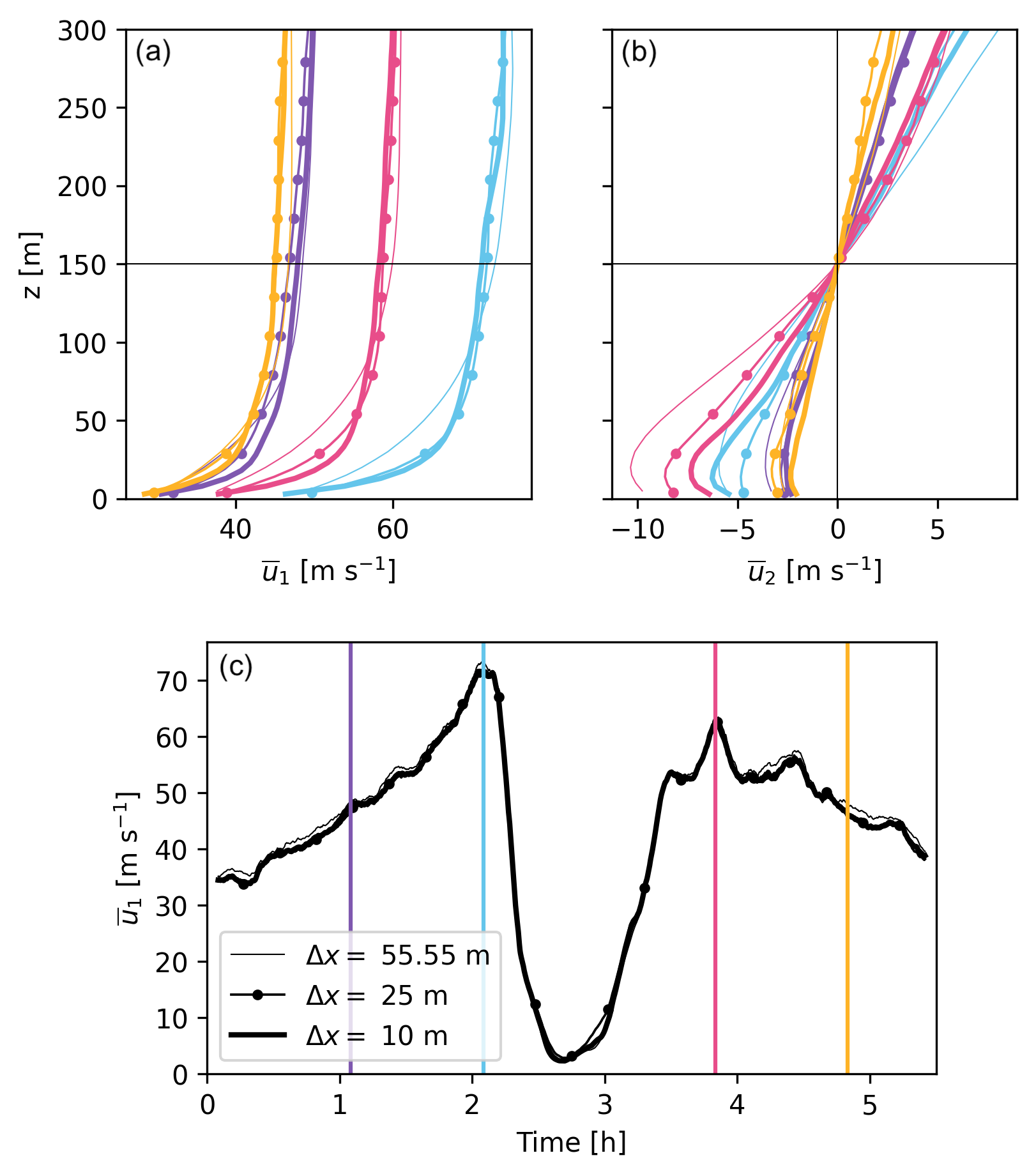}
\caption{Mean streamwise (a) and cross-stream (b) wind profiles. Panel (c) shows the temporal evolution of the mean horizontal wind speed at $z=150\,{\rm m}$ at the virtual turbine position. The vertical profiles in panels (a) and (b) are sampled at the four times shown in panel (c).}
\label{fig:meanWindProfiles}
\end{figure}

The mean hub-height winds that a turbine experiences during a hurricane passage vary significantly over time (Fig. \ref{fig:meanWindProfiles}). The streamwise mean wind speed increases across the entire rotor layer as the hurricane approaches the turbine position, decreases at the eye of the storm, increases again as the trailing edge of the eyewall approaches the turbine, and slows down over time afterward (Fig. \ref{fig:meanWindProfiles}). Due to asymmetries in the large-scale storm's structure, mean winds at the leading and trailing edges of the eyewall are different. The fastest winds occur at the leading edge of the eyewall (light-blue line in Fig. \ref{fig:meanWindProfiles}), followed by winds at the trailing edge of the eyewall (pink lines in Fig. \ref{fig:meanWindProfiles}). Mean streamwise winds are comparable $1\,{\rm h}$ before the leading edge of the eyewall reaches the turbine and $1\,{\rm h}$ after the trailing edge of the eyewall passes over the turbine.

\begin{figure}[htbp!] 
\centering \includegraphics[height=8cm]{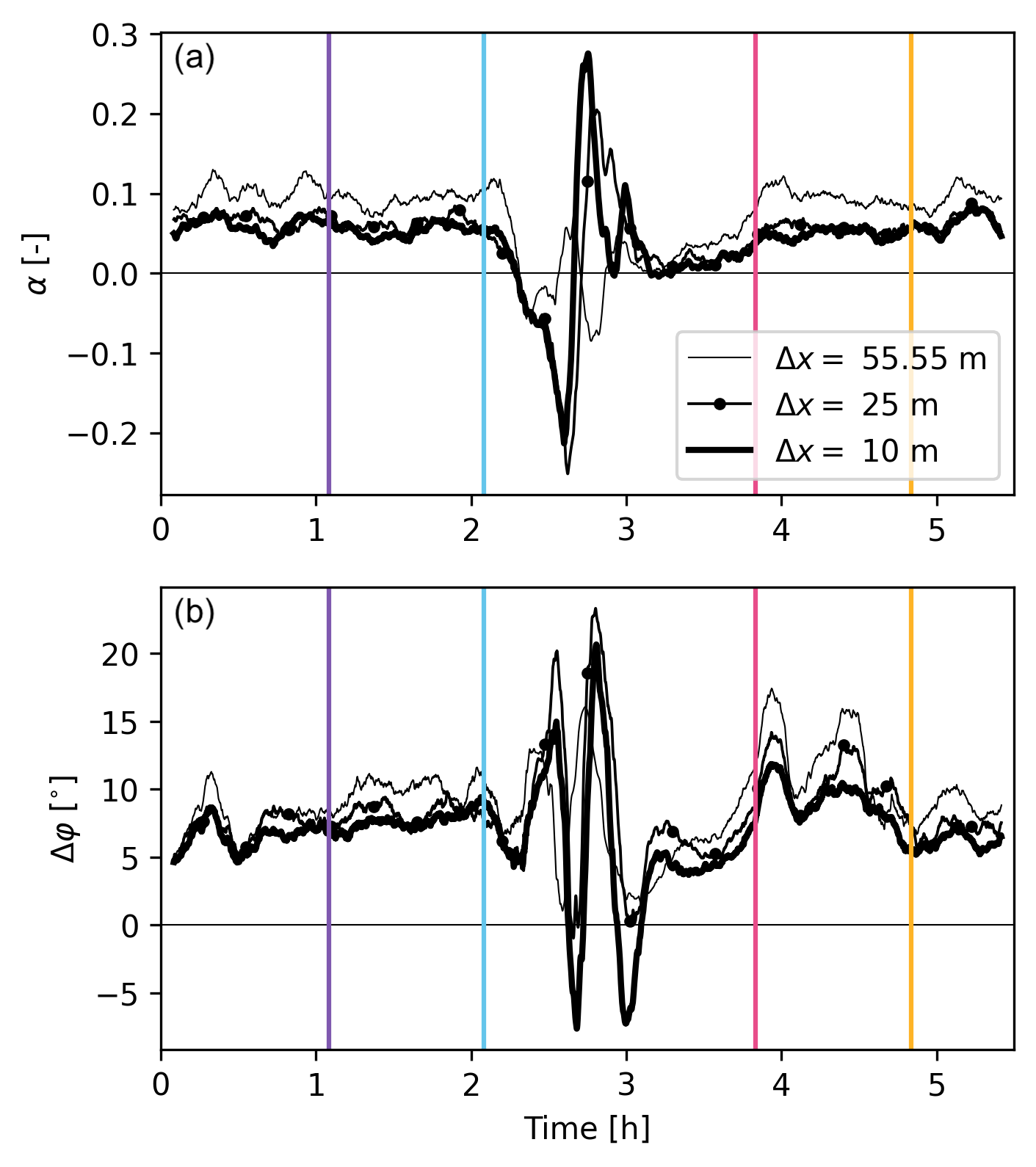}
\caption{Temporal evolution of the fit to the power-law wind exponent $\alpha$ (a) and of the vertical change in wind direction (b) across the turbine rotor layer.}
\label{fig:shear_veer}
\end{figure}

Like the mean wind speed magnitude changes over time, the vertical mean wind structure also changes (Fig. \ref{fig:shear_veer}). However, the mean wind profiles can differ before and after the eye of the storm moves over the turbine position, even if mean hub-height wind speed remains comparable. Mean wind speed shear hovers around $\alpha=0.057$ as the leading edge of the eyewall approaches the turbine. Inside the eye of the storm, the mean wind structure is not well characterized by the power-law wind profile due to quiescent wind conditions. After the trailing edge of the eyewall reaches the turbine position, wind speed shear increases from $\alpha=0.038$ to $\alpha=0.054$ far from the trailing edge of the eyewall. Mean wind direction shear also varies over time and remains positive everywhere, except at the eye of the storm, where wind direction may not be well defined. Wind veer increases from $\Delta\varphi=7^{\circ}$ far from the leading edge of the eyewall and reaches a local maximum of $\Delta\varphi=9^{\circ}$ as the leading edge of the eyewall impacts the turbine position. After the trailing edge of the eyewall passes the turbine position, mean wind veer fluctuates over time, reaching maximum values of $\Delta\varphi=11.9^{\circ}$.

The mean wind profile can vary with grid resolution. Vertical shear of the streamwise mean wind profile is different in the coarse- and fine-resolution simulations throughout the hurricane passage, especially near the surface, while differences between the intermediate- and fine-resolution LESs are small (Fig. \ref{fig:meanWindProfiles}a). The power-law wind exponent $\alpha$ for the coarse-resolution simulation is consistently larger than for the higher-resolution simulations, except when the turbine is directly in the eye of the storm (Fig. \ref{fig:shear_veer}a). For mean hub-height wind speeds exceeding $30\,{\rm m\,s^{-1}}$, the power-law exponent from the coarse-resolution LES is, on average, twice as large as for the fine-resolution LES. Conversely, the power-law exponent from the intermediate-resolution LES is only $10~\%$ larger, on average, than that of the fine-resolution simulation. Although the mean wind profiles can differ between the simulations, the mean wind speed at hub height remains nearly unchanged in all cases (Fig. \ref{fig:meanWindProfiles}c). Vertical shear of the cross-stream wind component is also usually smaller in the fine-resolution LES compared to the other simulations (Fig. \ref{fig:meanWindProfiles}b). The vertical change in wind direction across the rotor layer for the coarse-resolution simulation is consistently larger than for the fine-resolution simulation, except when the eye of the storm is directly over the turbine (Fig. \ref{fig:shear_veer}b). For mean hub-height wind speeds exceeding $30\,{\rm m\,s^{-1}}$, wind veer in the coarse- and intermediate-resolution LESs is, on average, $30~\%$ and $12~\%$ larger, respectively, compared to the fine-resolution LES. Mean cross-stream winds below hub height are usually larger in the coarser-resolution simulations while cross-stream winds above hub height remain comparable across all simulations (Fig. \ref{fig:meanWindProfiles}b). Therefore, differences in mean wind veer are for the most part due to differences in near-surface winds.

\subsection{Wind gusts}
Building on the analysis of mean wind profiles, we now focus on wind gusts, a key driver of turbine loading. To evaluate wind gusts, we apply a 3 s moving average to the wind data throughout the simulation time. We also consider the ratio of the short-term average (i.e., gust) wind speed to the mean wind speed, referred to as gust factors. Gust factors have historically been used in engineering applications to adjust measured wind speeds, averaged over a variety of timescales, to a consistent averaging time. Gust factor models, like the \cite{krayer_gust_1992} and \cite{durst_wind_1960} models, are commonly applied to adjust hurricane winds across timescales over land and for a height of $10$ $\rm m$ above the surface. However, the assumption of shear-driven neutrally stratified flow typically used to estimate gust factors above $10\,{\rm m}$ breaks down in the eyewall region, where convective structures enhance turbulence, likely modifying the gust factors above the surface \citep{krayer_gust_1992}. We calculate gust factors $(C_{g,\tau}(\mathbb{T}))$ as the ratio of the maximum $\tau$-averaged wind speed occurring within a $\mathbb{T}$ time period, and the $\mathbb{T}$-averaged wind speed within that same period (Eq. \eqref{eq:GF}). Here, we estimate gust factors at hub height $(z=150\,{\rm m})$ for $(\mathbb{T},\tau)=$ (10 min, 3 s). 

\begin{equation}
    C_{g,\tau}(\mathbb{T}) = \frac{{\max\limits_{t\,\in\,\mathbb{T}}}\left\{ \overline{U}_{\tau}(t)\right\}}{\overline{U}_{\mathbb{T}}(t)}
\label{eq:GF}
\end{equation}

\begin{figure}[htbp!] 
\centering \includegraphics[height=8cm]{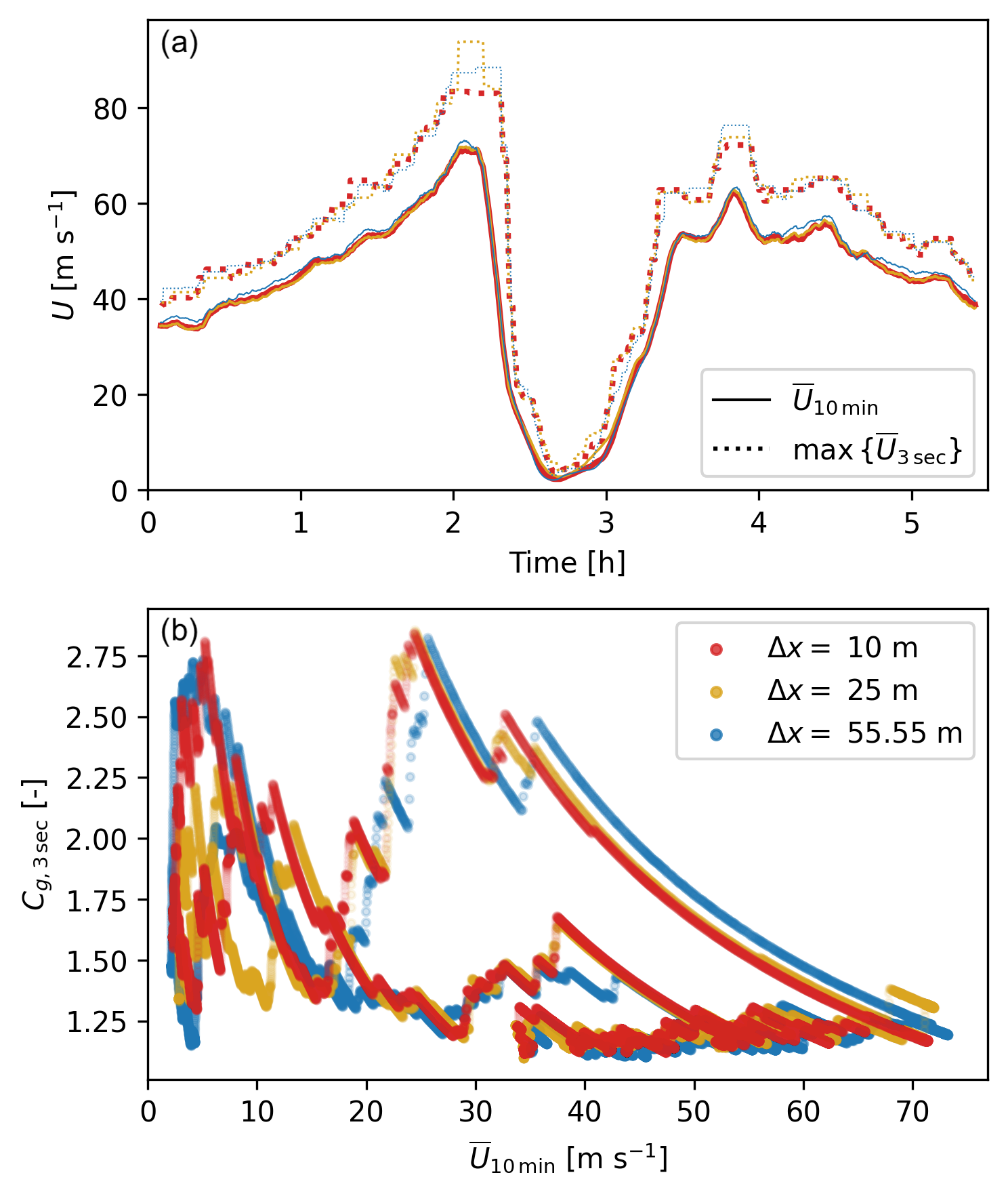}
\caption{Temporal evolution of mean wind speed and gusts at $z=150\,{\rm m}$ (a), color coded by the wind speed magnitude. Panel (b) shows the gust factor of 3 s wind gusts and 10 min averaged wind speeds at $z=150\,{\rm m}$.}
\label{fig:gusts_Cg}
\end{figure}

Like mean winds, hub-height wind gusts are largest at the storm's eyewall (Fig. \ref{fig:gusts_Cg}a). Maximum 3 s averaged wind gusts at hub height can exceed $80\,{\rm m\,s^{-1}}$ at the same time that 10 min averaged winds reach $70\,{\rm m\,s^{-1}}$. At the trailing edge of the eyewall, wind gusts can reach up to $72\,{\rm m\,s^{-1}}$ with mean winds of $62\,{\rm m\,s^{-1}}$. Because of the way they are defined, gust factors show a distinct behavior inside the eye of the storm and outside the eyewall. Outside the eyewall, both mean winds and gusts increase or decrease steadily over time. As a result, gust factors remain around $1.2$, on average. The largest gust factors occur as the leading edge of the eyewall impacts the virtual turbine location. At this time, the ${\max\limits_{t\,\in\,{10\,{\rm min}}}}\left\{ \overline{U}_{3\,{\rm sec}}(t)\right\}$ are largest but the 10 min average winds are decreasing in speed, producing $C_{g,\tau}(\mathbb{T})>1.4$ for wind speeds up to $\overline{U}_{10\,{\rm min}}\approx59\,{\rm m\, s^{-1}}$. 

\begin{figure}[htbp!] 
\centering \includegraphics[height=8cm]{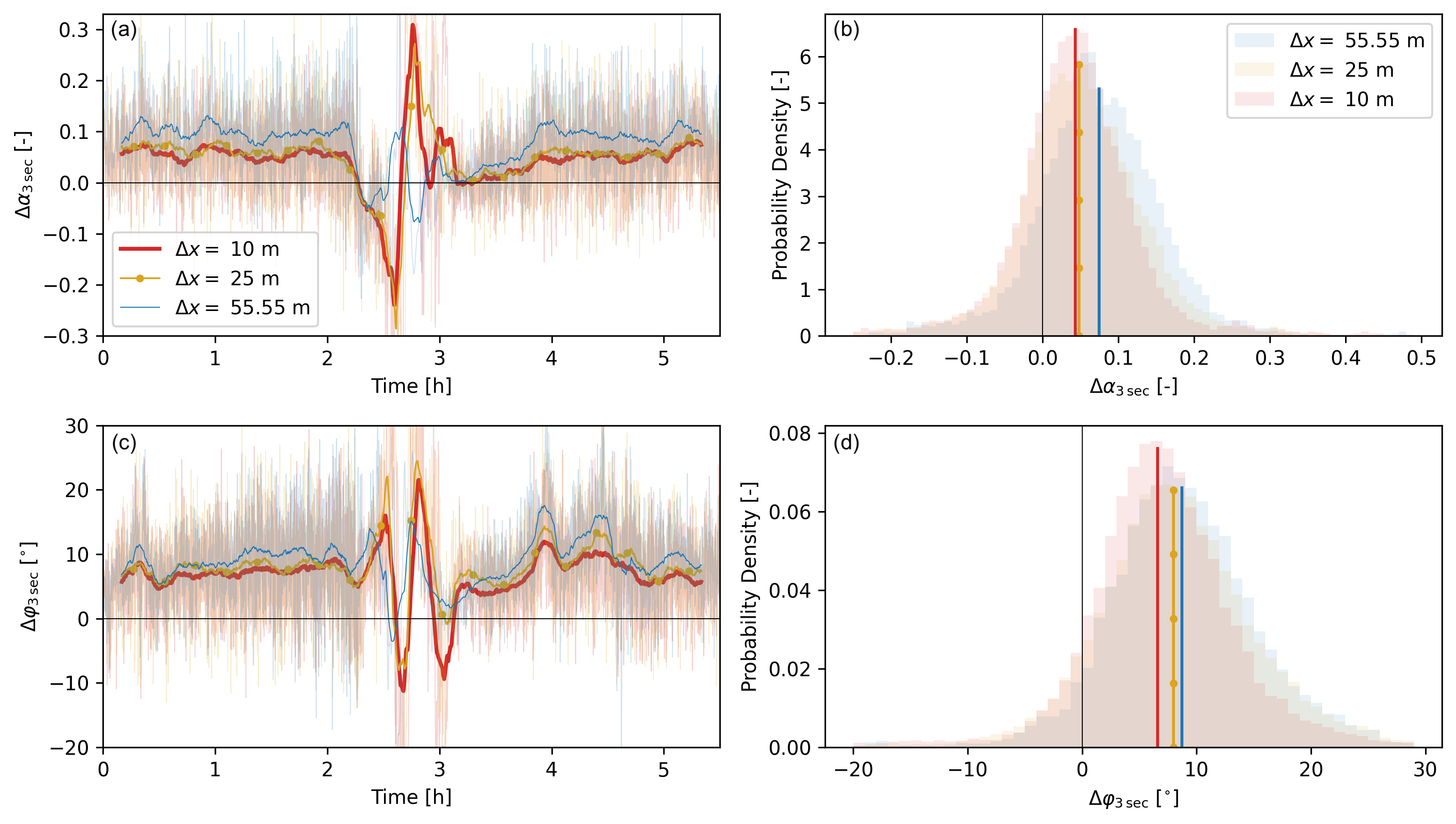}
\caption{Temporal evolution of vertical wind speed shear (a) and wind direction shear (b) calculated using 3 s averaged wind conditions. The probability distribution for wind speed and wind direction shear are shown in panels (c) and (d), respectively.}
\label{fig:shear_veer_gust}
\end{figure}

Wind gusts show large variability in wind speed and direction shear throughout the hurricane passage (Fig. \ref{fig:shear_veer_gust}). Wind speed and direction shear in wind gusts are, on average, $\alpha_{3\,{\rm sec}}=0.049$ and $\Delta \varphi_{3\,{\rm sec}}=7.1\,^{\circ}$, respectively, for $\overline{U}_{10\,\rm min}>30\,{\rm m\,s^{-1}}$. For mean hub-height wind speeds exceeding $30\,\rm m\,s^{-1}$, negative wind speed shear occurs about $20~\%$ of the time and wind backing about $8~\%$ of the time. Wind speed shear remains around $\alpha_{3\,{\rm sec}}=0.057$ (with a standard deviation of $\sigma_{\alpha}=0.051$) before the leading edge of the eyewall reaches the turbine position. After the trailing edge of the eyewall passes over the turbine, wind speed shear increases from about $\alpha_{3\,{\rm sec}}=0.039$ $(\sigma_{\alpha}=0.054)$ up to $\alpha_{3\,{\rm sec}}=0.082$ $(\sigma_{\alpha}=0.05)$ far from the eyewall. Before the leading edge of the eyewall reaches the turbine, wind direction shear in gusts is, on average, $\Delta \varphi_{3\,{\rm sec}}=5^{\circ}$ with a standard deviation of $\sigma_{\Delta \varphi}=6.86^{\circ}$. Wind direction shear displays more variability after the eye moves over the turbine, where the average veer can reach $\Delta \varphi_{3\,{\rm sec}}=10^{\circ}$ $(\sigma_{\Delta \varphi}=5.47^{\circ})$ at the trailing edge of the eyewall and decreases to about $\Delta \varphi_{3\,{\rm sec}}=5^{\circ}$ $(\sigma_{\Delta \varphi}=6.68^{\circ})$ afterward.

Although mean winds at hub height are comparable between the coarse-, intermediate-, and fine-resolution simulations, wind gusts can differ, especially at the eyewall of the storm (Fig. \ref{fig:gusts_Cg}). All simulations show that the fastest 3 s hub-height wind gusts occur when the leading edge of the eyewall moves over the virtual turbine position (Fig. \ref{fig:gusts_Cg}a). However, the coarse- and intermediate-resolution simulations overestimate the most extreme wind gusts at the eyewall. Wind gusts in the leading and trailing edge of the eyewall are $6~\%$ and $5~\%$ faster, respectively, in the coarse-resolution simulation than in the fine-resolution simulation. While the intermediate-resolution LES does not overestimate gusts in the trailing edge of the eyewall, hub-height wind gusts in the leading edge are $12~\%$ faster than in the fine-resolution simulation. Because wind gusts can be faster in the coarser-resolution simulations but the 10 min average winds are nearly equal, the gust factor is also larger in the coarser LESs (Fig. \ref{fig:gusts_Cg}b). The gust factor can exceed $1.4$ for wind speeds up to $\overline{U}_{10\,{\rm min}}\approx63\,{\rm m\, s^{-1}}$ in the coarse-resolution simulation, and $\overline{U}_{10\,{\rm min}}\approx59\,{\rm m\, s^{-1}}$ in the intermediate- and fine-resolution LESs.

Like the mean wind structure across the rotor layer can differ between the simulations, the vertical structure of wind gusts can also vary (Fig. \ref{fig:shear_veer_gust}). Wind speed shear in gusts is generally larger in the coarse-resolution LES compared to the intermediate- and fine-resolution simulations (Fig. \ref{fig:shear_veer_gust}a,b). The power-law exponent for the 3 s averaged wind speed profiles is, on average, $0.074$, $0.048$, and $0.043$ for the coarse-, intermediate-, and fine-resolution simulations, respectively. Furthermore, for times when the mean hub-height wind speed exceeds $30\,{\rm m\,s^{-1}}$, the power-law exponent $\alpha_{3\,{\rm sec}}$ is $0.088$, $0.055$, and $0.048$ for the coarse-, intermediate-, and fine-resolution simulations, respectively. The vertical change in 3 s averaged wind direction is also larger in the coarse-resolution simulation (Fig. \ref{fig:shear_veer_gust}c,d). Vertical wind direction shear is $8.8^{\circ}$, $8.1^{\circ}$, and $6.7^{\circ}$, on average, for the coarse-, intermediate-, and fine-resolution simulations, respectively. Excluding the eye of the storm, wind direction shear is $9.5^{\circ}$, $8.0^{\circ}$, and $7.1^{\circ}$, on average, for the coarse-, intermediate-, and fine-resolution simulations, respectively.

\subsection{Turbulence statistics}
Having characterized the vertical structure of mean winds and gusts, we now study the nature of turbulence throughout the hurricane passage at turbine-relevant heights. 
\subsubsection{Velocity variances}
As a first step in characterizing turbulence, we examine the turbulence content of the flow, defined by the magnitude of the velocity variances. The velocity variances are estimated from the fluctuation velocities using a 10 min moving average. We examine the variance $\overline{u'_iu'_i}$ of each velocity component individually to understand limitations within each grid. Note that we only consider the resolved component of the Reynolds stress tensor in the LES because the resolved velocity field is used as direct inflow to the aeroelastic simulations, without including contributions from the modeled stress components. 

\begin{figure}[htbp!] 
\centering \includegraphics[width=8cm]{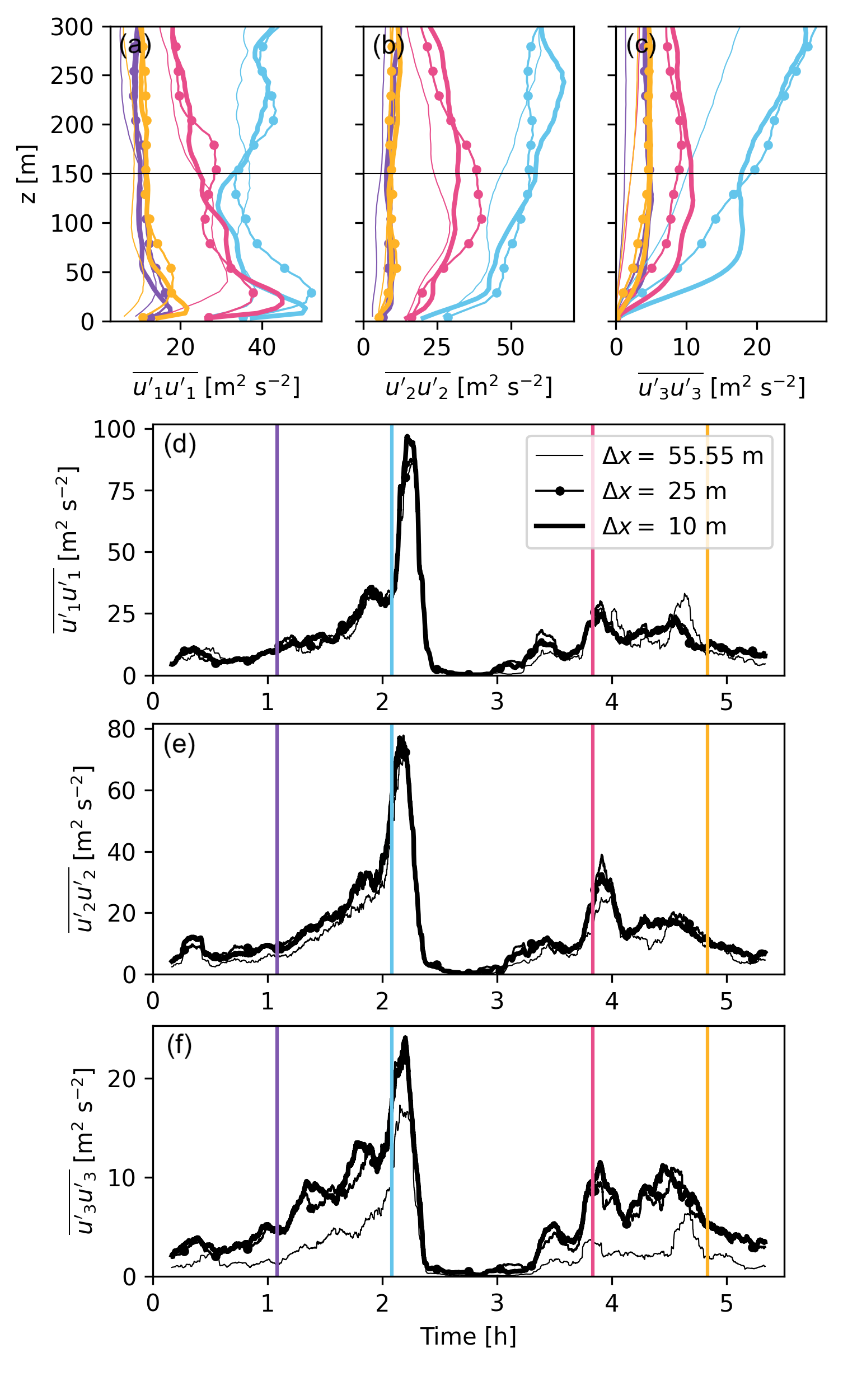}
\caption{Vertical profiles of the streamwise (a), cross-stream (b), and vertical (c) velocity variances in the lowest $300\,{\rm m}$ of the boundary layer. The temporal evolution of the velocity variances at $z=150\,{\rm m}$ at the virtual turbine position are shown in panels (d--f). The vertical profiles in panels (a--c) are sampled at the four times shown in panels (d--f).}
\label{fig:turbProfiles}
\end{figure}

The turbulence content of the flow changes dramatically over time as the hurricane passes over the virtual turbine position (Fig. \ref{fig:turbProfiles}). The hub-height velocity variances increase from $(\overline{u'_1u'_1},\overline{u'_2u'_2},\overline{u'_3u'_3})=(5,5,3)\,{\rm m^{2}\,s^{-2}}$ far from the leading edge of the eyewall up to $(\overline{u'_1u'_1},\overline{u'_2u'_2},\overline{u'_3u'_3})=(97,77,24)\,{\rm m^{2}\,s^{-2}}$ immediately after the leading edge of the eyewall moves over the turbine position. Surprisingly, the peak turbulence content of the flow (corresponding to $(\overline{u'_1u'_1},\overline{u'_2u'_2},\overline{u'_3u'_3})=(97,77,24)\,{\rm m^{2}\,s^{-2}}$) occurs immediately after the eyewall moves over the turbine and not when mean hub-height winds are fastest (corresponding to $ (\overline{u'_1u'_1},\overline{u'_2u'_2},\overline{u'_3u'_3})=(33,58,18)\,{\rm m^{2}\,s^{-2}}$). The magnitude of the velocity variances increases after the eye of the storm passes over the virtual turbine position. However, the velocity variances are smaller in the trailing edge of the eyewall compared to the leading edge. The maximum in the streamwise, cross-stream, and vertical velocity variances at hub height in the trailing edge of the eyewall is $25\,{\rm m^{2}\,s^{-2}}$, $32\,{\rm m^{2}\,s^{-2}}$, and $12\,{\rm m^{2}\,s^{-2}}$, respectively. Moreover, the maximum in the velocity variances after the eye moves over the turbine occurs at the same time as the maximum in mean hub-height wind speed. 

The resolved velocity variances are almost always larger in the turbine-scale simulations compared to the coarse-resolution LES (Fig. \ref{fig:turbProfiles}), as expected. The hub-height streamwise, cross-stream, and vertical velocity variances are larger in the fine-resolution LES than in the coarse-resolution LES $70~\%$, $90~\%$, and $99~\%$ of the time, respectively (Fig. \ref{fig:turbProfiles}c--d). Moreover, for mean hub-height wind speeds exceeding $30\,{\rm m\,s^{-1}}$, the hub-height streamwise, cross-stream, and vertical velocity variances are $27~\%$, $29~\%$, and $60~\%$ larger in the fine LES than in the coarse LES, on average. Although the velocity variances are also frequently larger in the fine LES than in the intermediate LES, differences in the velocity variances are smaller than $17~\%$, on average. Differences in velocity variances persist across the turbine rotor layer (Fig. \ref{fig:turbProfiles}a,b). The streamwise and cross-stream velocity variances across the rotor layer are, on average, $18~\%$ larger in the fine-resolution LES than in the coarse LES, while the vertical velocity variance is $66~\%$ larger. On average, differences in the streamwise, cross-stream, and vertical velocity variances across the rotor layer between the intermediate and fine LESs are $10.5~\%$, $12~\%$, and $16~\%$, respectively. 

\subsubsection{Integral length scales \label{sec:L}}
Also important for wind turbine loads are the size of the dominant eddies in the flow \citep{stanislawski_effect_2023}. We estimate the integral length scale of each velocity component $(u_i)$ along the cross-stream direction $(y^*)$. The two-point correlation function of time series velocity data at the virtual turbine position is defined by Eq. \eqref{eq:corrFunc}, where $\delta$ is the distance along the cross-stream direction (i.e., along the span of the turbine) between the sampling points. The integral length scale $L_{u_i}^{y^*}$ is then estimated following Eq. \eqref{eq:L}, where $\delta_{\rm max}$ is the distance where the two-point correlation becomes negative. 

\begin{eqnarray}
\rho_{u_i u_i}^{y^*}(\delta) &=& \frac{\overline{u'_i(y^*)\,u'_i(y^*+\delta)}}{\overline{u'_i(y^*)^2}} \label{eq:corrFunc} \\
L_{u_i}^{y^*} &=& \int_0^{\delta_{\rm max}} \rho_{u_i u_i}^{y^*}(\delta) \, d\delta,  \label{eq:L} 
\end{eqnarray}

\begin{figure}[htbp!] 
\centering \includegraphics[height=10cm]{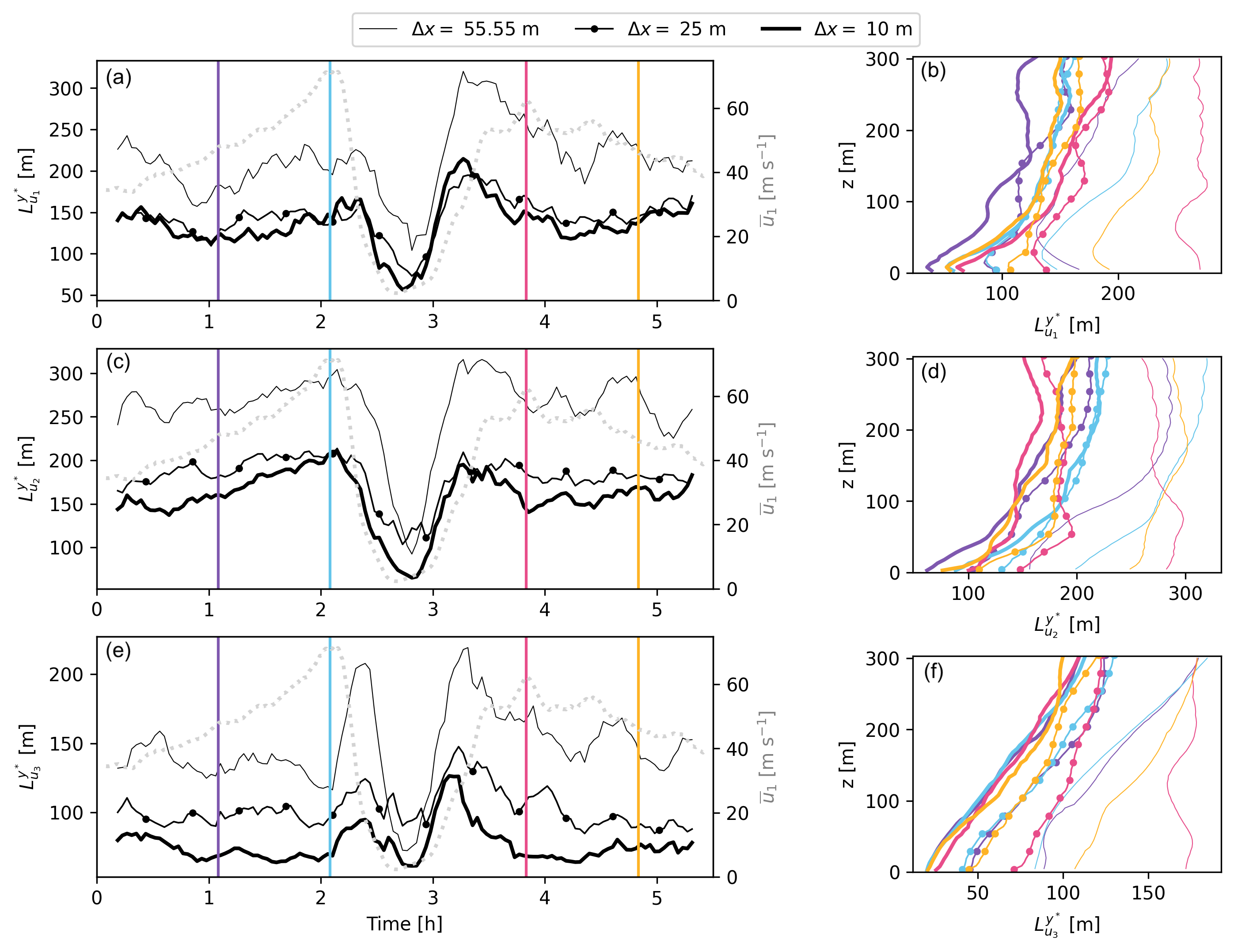}
\caption{Temporal evolution of the integral length scale of the streamwise (a), cross-stream (c), and vertical (e) velocity components at hub height and along the cross-stream direction. The vertical profiles of the streamwise (b), cross-stream (d), and vertical (f) wind components are shown for the times four of interest. The secondary $y$-axes in panels (a,c,e) show the temporal evolution of the mean streamwise wind speed at hub height for reference (gray dotted line).}
\label{fig:L_crsstrm}
\end{figure}

Changes in the hub-height integral length scales of the flow remain small as the hurricane passes over the virtual turbine position, except for very slow wind speeds (Fig. \ref{fig:L_crsstrm}a,c,e). The integral length scale of the streamwise wind at hub height remains nearly unchanged as the leading edge of the eyewall approaches the turbine location $(L_{u_1}^{y^*}\approx132\,{\rm m})$ and after the trailing edge of the eyewall passes by $(L_{u_1}^{y^*}\approx139\,{\rm m})$. Similarly, the hub-height integral length scale of the vertical wind is about $L_{u_3}^{y^*}\approx73\,{\rm m}$ before and after the eyewall hit. Conversely, the integral length scale of the cross-stream wind increases steadily from $L_{u_2}^{y^*}=150\,{\rm m}$ far from the leading edge of the eyewall up to $L_{u_2}^{y^*}=210\,{\rm m}$ as the leading edge of the eyewall reaches the turbine position. After the trailing edge of the eyewall moves over the turbine, the integral length scale of the cross-stream wind remains nearly unchanged at $L_{u_2}^{y^*}\approx160\,{\rm m}$.

The temporal variability in integral length scales is much smaller than the variability with height (Fig. \ref{fig:L_crsstrm}b,d,f). The vertical profiles of $L_{u_3}^{y^*}$ at the four times of interest indicate that the size of the dominant eddies in the vertical velocity field scales linearly with height with minimal temporal variations (Fig. \ref{fig:L_crsstrm}f). The integral length scales of the streamwise and cross-stream wind components do not scale linearly with height. However, $L_{u_1}^{y^*}$ and $L_{u_2}^{y^*}$ fluctuate by only tens of meters over time but change by hundreds of meters across the vertical profile (Fig. \ref{fig:L_crsstrm}b,d). Small temporal variability in the size of the dominant eddies in the lowest $300\,{\rm m}$ for wind speeds above $30\,{\rm m\,s^{-1}}$ may be due to proximity to the surface, where the size of the eddies scales with height rather than with the larger-scale structure of the storm. Because the flow is highly shear-driven and because the surface layer can extend to the top of the turbine rotor layer, local shear production dominates turbulence dynamics, and the limiting length scale continues to be geometric (distance to the surface).

The integral length scales of the flow can vary significantly in the eye of the storm, where mean wind speeds suddenly slow down (Fig. \ref{fig:L_crsstrm}a,c,e). Within the storm's eye, mean wind speeds at hub height slow down to $2\,{\rm m\,s^{-1}}$ and turbulence is almost non-existent (Fig. \ref{fig:turbProfiles}c,d), resulting in local minima for the integral length scale of all three wind components. The sizes of the dominant eddies in the flow appear largest immediately after the leading edge of the eyewall reaches the turbine position and just before the trailing edge of the eyewall moves over the turbine position (Fig. \ref{fig:L_crsstrm}), just like the velocity variances (Fig. \ref{fig:turbProfiles}). Note, however, that the integral length scales of the flow immediately after the leading edge of the eyewall passes over the turbine (or immediately before the trailing edge of the eyewall) may not be well defined. Immediately after the leading edge of the eyewall (and immediately before the trailing edge of the eyewall) moves over the turbine, the wind field is highly non-stationary (mean hub-height winds can vary by more than $40\,{\rm m\,s^{-1}}$ within a 10 min period). Consequently, the integral length scale of the flow will also vary within the same 10 min period. Because we are estimating the integral length scale over 10 min overlapping segments, it is likely that large changes in the integral length scale near the storm's eyewall are affected by highly non-stationary conditions.

The integral length scale of each velocity component at hub height is typically larger in the coarse- and intermediate-resolution simulations than in the fine LES, especially for the vertical wind component (Fig. \ref{fig:L_crsstrm}a,c,e). The integral length scale of the streamwise velocity is $50~\%$ larger, on average, in the coarse-resolution simulation than in the fine LES. The integral length scale of the streamwise velocity is, on average, 8~\% larger in the intermediate LES than in the fine LES. Differences in the integral length scale for the cross-stream and vertical velocity components are more pronounced. The dominant eddies at hub height are $63~\%$ and $94~\%$ larger in the coarse LES than in the fine LES for the cross-stream and vertical velocity components, respectively. Length scales of the cross-stream and vertical wind components in the intermediate LES are $17~\%$ and $35~\%$ larger, on average, than in the fine LES, respectively. The size of the dominant eddies in the flow is also generally larger across all heights (Fig. \ref{fig:L_crsstrm}b,d,f). Moreover, contrary to the fine-resolution simulation results, the integral length scales of the flow across the lowest $300\,{\rm m}$ can change significantly over time in the coarse-resolution simulation. Large temporal variations in $L_{u_i}^{y^*}$ are likely caused by limitations in resolving turbulence with the coarser grid.  \\

\subsubsection{Turbulence representation across scales} \label{sec:turbSpctr}
After quantifying the turbulence content of the flow and the size of the dominant eddies, we now characterize the distribution of energy across scales and the size of the smallest eddies resolved by each grid. We use the energy spectra of the hub-height streamwise and vertical velocity components to quantify the range of eddies resolved by each simulation. The energy spectra of the cross-stream velocity are not included in the analysis, as they resemble that of the streamwise wind. To evaluate how turbulent energy is captured across scales, we estimate the range of frequencies captured by each grid within the inertial subrange $(S(f)\sim f^{-5/3})$ of the flow at turbine heights. The range of scales resolved within the inertial subrange in each simulation are estimated by fitting the equilibrium Kolmogorov spectrum of the form $S_{ii}(k)=C_i \varepsilon^{2/3}k^{-5/3}$, where $C_1\approx0.5$ and $C_{2,3}\approx0.65$ \citep{pope_turbulent_2015}, to the one-dimensional spectrum of each velocity component at hub height. To remain consistent with the forthcoming analysis, we express the Kolmogorov spectrum in terms of frequency by employing Taylor's frozen turbulence hypothesis $(k=2\pi f/\overline{u}_1)$, as shown in Eq. \eqref{eq:spctrum}. 

\begin{equation}
    S_{ii}(k)=\left ( \frac{2\pi}{\overline{u}_1} \right)^{-5/3} C_i \varepsilon^{2/3}f^{-5/3} \label{eq:spctrum}
\end{equation}

We estimate the range of frequencies resolved by each grid within the inertial subrange by calculating a local spectral slope with a moving window across frequencies. We fit the energy spectra to the Kolmogorov spectrum for frequencies between the spectral peak (i.e., the maximum in $fS_{ii}$) and the cutoff frequency of the simulations $(f_{\rm cutoff}=\overline{u}_1k_{\rm cutoff}/2\pi)$. The cutoff frequency is a function of the local advective velocity in the flow, here assumed to be equal to the mean streamwise velocity, and the effective resolution of the model, assumed to be $k_{\rm cutoff}\approx 1/7\Delta x$ \citep{skamarock_evaluating_2004}, as $f_{\rm cutoff}\approx\overline{u}_1/14\pi\Delta x$. Note that because the grid spacing of each simulation remains fixed but the streamwise velocity changes throughout the hurricane passage, the range of frequencies resolved by each grid are expected to change. Like with the previous analysis, we first consider the results from the fine-resolution simulation and then compare the distribution of energy across scales between different grid spacing LESs.

\begin{figure}[htbp!] 
\centering \includegraphics[height=7cm]{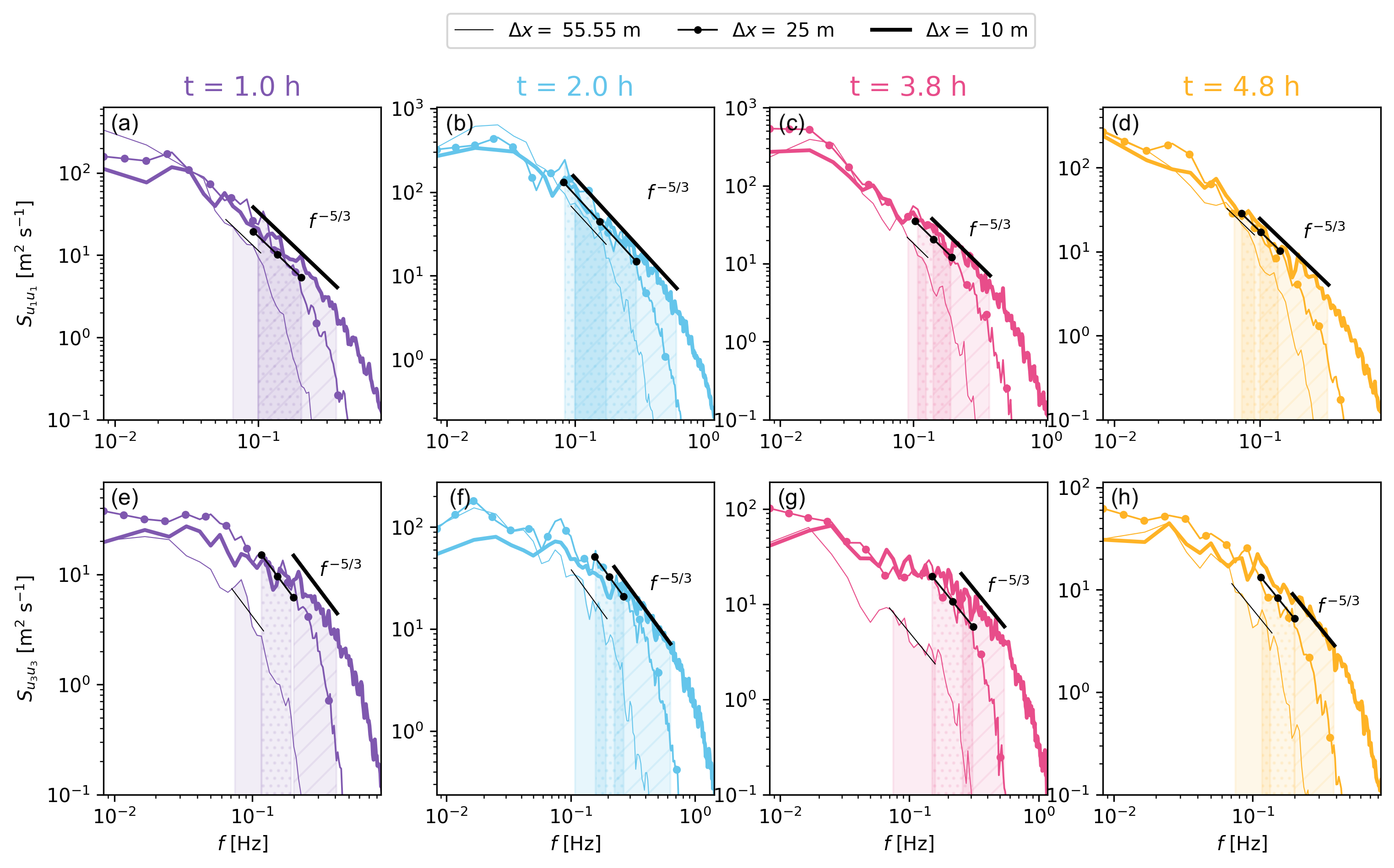}
\caption{Energy spectra of the streamwise (a--d), and vertical (e--h) wind components. Results are shown for the four times of interest. The hatched areas under each line represent the range of frequencies that characterize the inertial subrange (i.e, $S_{ii}\sim f^{-5/3}$) for each grid. }
\label{fig:spectra_ui}
\end{figure}

The onset of the inertial subrange of the flow changes between the horizontal and vertical wind components (Fig. \ref{fig:spectra_ui}). Because eddies in the vertical velocity field are smaller than in the streamwise velocity field (Fig. \ref{fig:L_crsstrm}), the frequencies associated with the inertial subrange of the flow are expected to be larger for $u_3$ compared to $u_1$. As a result, the inertial subrange generally begins at around $f\approx0.09$--$0.14\,{\rm Hz}$ for $u_1$, and at around $f\approx0.2$--$0.27\,{\rm Hz}$ for $u_3$. In addition, just as the integral length scales of the flow change minimally throughout the hurricane passage (Sect. \ref{sec:L}), the onset of the inertial subrange also remains near constant in time. However, the maximum frequency $(f_{\rm max}\sim \overline{u}_1/\Delta x)$ resolved by the grid varies because the advective velocity of the flow changes as the hurricane moves over the virtual turbine position, but the grid spacing of the simulation remains fixed. As a result, a wider portion of the inertial subrange is being resolved at the eyewall of the storm (Fig. \ref{fig:spectra_ui}b,f) compared to far from the eyewall (Fig. \ref{fig:spectra_ui}d,h).  

The distribution of energy across scales can differ between the coarse-, intermediate-, and fine-resolution simulations (Fig. \ref{fig:spectra_ui}). In general for the streamwise wind, the coarser-resolution simulations contain more energy at low frequencies (i.e., large-scale features in the flow), whereas the fine-resolution simulation contains more energy at high frequencies (i.e., small-scale features in the flow). For the streamwise wind, the coarse- and intermediate-resolution simulations contain $37~\%$ and $33~\%$ more variance than the fine-resolution simulation, on average, for frequencies smaller than $0.06\,{\rm Hz}$. For the vertical wind component, the coarse LES generally underestimates the variance contained at low frequencies, whereas the intermediate-resolution LES always overestimates it. On average, normalized absolute-value differences in the $u_3$ variance contained within the large-scale flow are $31~\%$ for the coarse and fine LES. The intermediate-resolution simulation contains $51~\%$ more variance than the fine LES for frequencies smaller than $0.06\,{\rm Hz}$, on average. Differences in energy content at high frequencies are expected due to differences in the grid resolution of the simulations. However, higher energy content in the flow at low frequencies indicates that the energy transfer between large and small scales differs between the simulations.

The inertial subrange of the flow at hub height typically falls within frequencies that are not resolved by the coarse grid (Fig. \ref{fig:spectra_ui}). Based on the energy spectra from the fine-resolution simulation, the inertial subrange of the streamwise velocity typically starts around $f\gtrapprox0.1\,{\rm Hz}$. However, the energy spectra of the streamwise velocity from the coarse-resolution simulation starts deviating from that of the high-resolution simulation around $f\approx{7\times10^{-2}\,\rm Hz}$ far from the eyewall, and around $f\approx{0.2\,\rm Hz}$ at the eyewall (Fig. \ref{fig:spectra_ui}a--d). Differences in the energy spectra of the vertical velocity between both simulations are even more pronounced. The coarse-resolution simulation cannot resolve the inertial subrange of the vertical velocity field throughout the entire hurricane passage (Fig. \ref{fig:spectra_ui}e--h). Because the inertial subrange is generally not well resolved by the coarse grid, the variance contained within those frequencies is underestimated compared to the fine-resolution simulation. This means that the coarse-resolution simulation is not capable of adequately resolving turbulent eddies with characteristic frequencies higher than $f\gtrapprox0.1\,{\rm Hz}$ throughout the entire hurricane passage at hub height. As a result, the natural frequencies of the turbine's tower $(f_{n,t}=0.17\,{\rm Hz})$ and blades $(f_{n,b}=0.56\,{\rm Hz})$ will likely not be excited by the coarse LES inflow.

The intermediate-resolution simulation partially captures the inertial subrange at hub height near the hurricane eyewall (Fig. \ref{fig:spectra_ui}). For the streamwise velocity, the intermediate LES resolves the onset of the inertial subrange near the leading and trailing edges of the eyewall (Fig. \ref{fig:spectra_ui}b,c), but predicts an earlier onset farther from the eyewall (Fig. \ref{fig:spectra_ui}a,d). For the vertical velocity, the onset of the inertial subrange occurs at $f \gtrapprox 0.15\,\mathrm{Hz}$ in the intermediate LES, compared to $f \gtrapprox 0.2\,\mathrm{Hz}$ in the fine LES. Because the intermediate LES generally fails to capture the vertical-velocity inertial subrange, energy transfer between large and small scales is inadequately represented, resulting in excess variance at low frequencies. While the intermediate LES accurately captures the streamwise energy near the tower’s natural frequency, it does not resolve the natural frequencies of the blades. Moreover, its poor representation of the vertical-velocity energy distribution may prevent full excitation of these frequencies in the turbine response.

\section{Wind turbine response \label{sec:windturb_response}}
After studying the velocity fields in the lowest portion of the boundary layer during the hurricane passage and their differences with grid resolution, we now consider the wind turbine's response to the different inflow wind fields. We investigate the wind turbine's response to the hurricane passage in terms of the tower-base and blade-root bending moments obtained from the aeroelastic simulations, which are representative of the turbine response \citep{griffith_structural_2009,anderson_development_2022}. We estimate the magnitude of the tower-base bending moment as $|M_t|=\sqrt{M_x^2+M_y^2}$, where $x$ and $y$ are in a Cartesian coordinate system. The magnitude of the blade-root moment for the $i$-th blade in the turbine is defined in a blade-aligned coordinate system as $|M_{b,i}|=\sqrt{M_{x_{b,i}}^2+M_{y_{b,i}}^2}$, where $y_{b,i}$ points toward the trailing edge of the blade, $z_{b,i}$ points along the pitch axis toward the tip of the blade, and $x_{b,i}$ is orthogonal to the $y_{b,i}$ and $z_{b,i}$ axes \citep{jonkman_fast_2005}. Here, we examine the 10 min averaged loads, as well as the maximum instantaneous loads occurring over 10 min moving windows. And, as with the wind fields, we first consider the turbine's response under the fine-resolution inflow, and then compare it with the turbine's response from the coarse- and intermediate-resolution inflows.

\begin{figure}[htbp!] 
\centering \includegraphics[height=9cm]{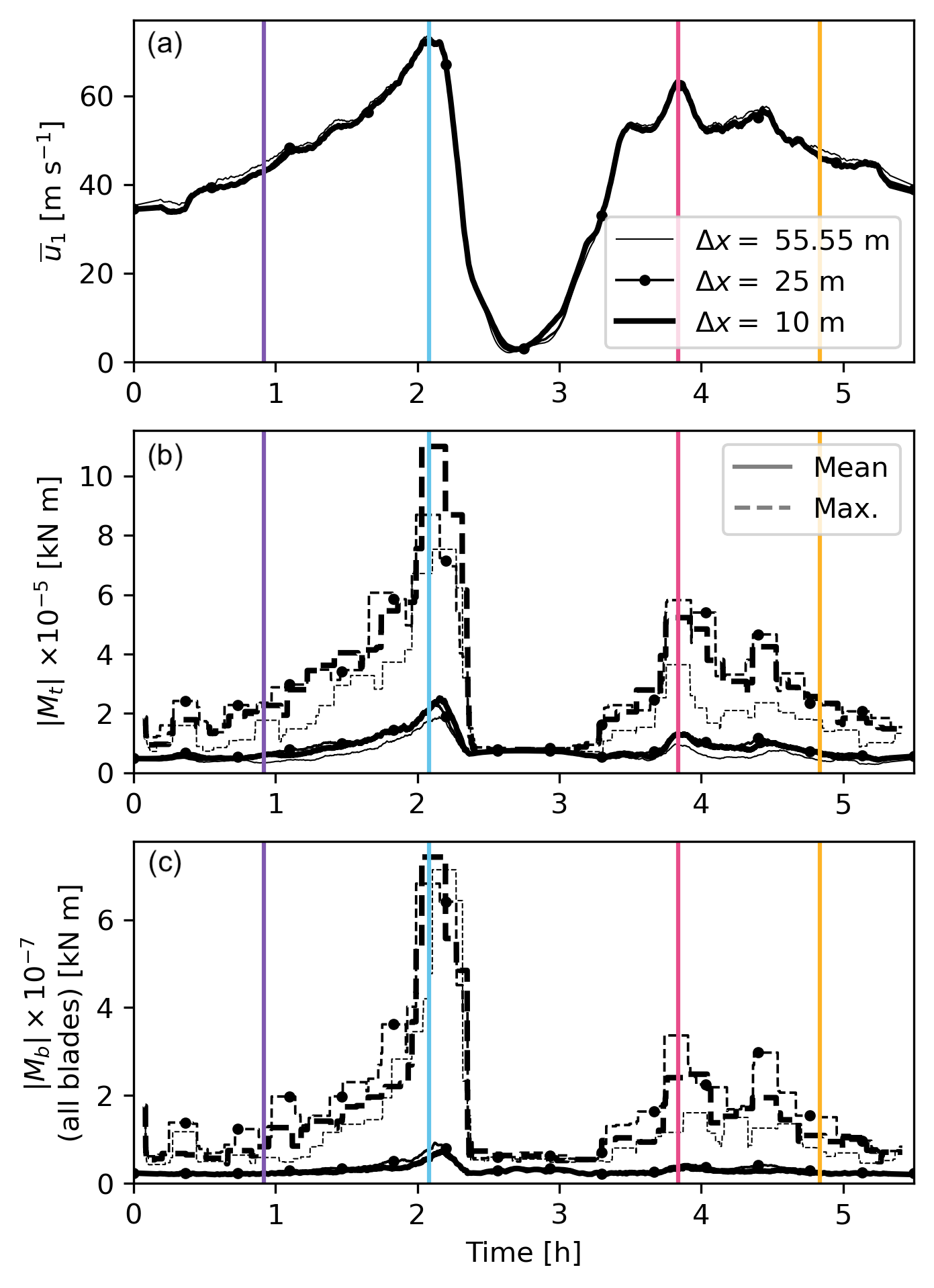}
\caption{Temporal evolution of hub-height wind speed (a), tower-base moment (b), and blade-root moment (c) throughout the hurricane passage. Mean values are shown using solid lines, and the moving maximum is represented by dashed lines. }
\label{fig:loads_turb_time}
\end{figure}

The blade-root and tower-base moments are largest near the eyewall of the storm and smallest in the eye (Fig. \ref{fig:loads_turb_time}), as expected. The maximum mean and instantaneous loads on both the tower base and blade root occur as the leading edge of the eyewall passes over the turbine. Notably, the maximum moments on the turbine structure are observed 2 min before the peak in mean wind speed. Wind turbine loads increase again as the trailing edge of the eyewall impacts the turbine, but to a lesser degree. For the trailing edge of the eyewall, the maximum mean and instantaneous loads also occur 1 min after the maximum mean hub-height wind speed is observed.

Tower-base moments are almost always underestimated when using inflow from the coarse-resolution simulation (Fig. \ref{fig:loads_turb_time}b). The maximum and mean tower-base moments obtained from the fine-resolution inflow are larger $92~\%$ and $95~\%$ of the time, respectively, compared to those obtained from the coarse-resolution inflow. The mean and maximum tower-base moments are, on average, $21.5~\%$ and $28.5~\%$ smaller when using the coarse-resolution inflow compared to the fine-resolution inflow. Differences in loads are even larger when only considering wind conditions outside the eye of the storm. For mean hub-height wind speeds above $30\,{\rm m\,s^{-1}}$, the mean and maximum tower-base moments are, on average, $25.6~\%$ and $32.5~\%$ smaller when using the coarse-resolution inflow compared to the fine-resolution inflow. Interestingly, the largest differences between the tower-base moments obtained from the coarse- and fine-resolution inflow occur after the trailing edge of the eyewall moves over the turbine position (at $t\approx4\,{\rm h}$). Nevertheless, the maximum tower-base moment experienced by the turbine throughout the entire hurricane passage is $31~\%$ larger when using the fine-resolution inflow.

The tower-base bending moments obtained from the intermediate LES inflow are similar to those obtained from the fine-resolution inflow (Fig. \ref{fig:loads_turb_time}b). The mean absolute difference in the maximum tower-base loads between the fine- and intermediate-resolution inflow is $15~\%$, on average. The largest differences occur as the leading edge of the eyewall impacts the turbine position, where the maximum bending moment is $21~\%$ larger for the fine LES inflow conditions. Differences in the mean loads are smaller. The mean absolute deviation in the average tower-base bending moments is $9~\%$, on average, throughout the hurricane passage.

Differences in the mean and maximum blade-root moments across inflow conditions are generally smaller than those observed for the tower-base moments (Fig. \ref{fig:loads_turb_time}c). The maximum blade-root moments obtained from the fine-resolution inflow are larger $88~\%$ of the time compared to those obtained from the coarse-resolution inflow. Nevertheless, the maximum blade-root moments are about $17.2~\%$ smaller when using the coarse-resolution inflow, on average. Differences in the mean values are smaller than $5~\%$, on average. Surprisingly, the mean and maximum blade-root moments are, on average, $16~\%$ and $27~\%$ larger for the intermediate-resolution inflow compared to the fine-resolution inflow. Nevertheless, the maximum blade-root moment experienced by the turbine throughout the entire hurricane passage is still underestimated when using the coarse- and intermediate-resolution inflows. The maximum blade-root loads from the fine LES inflow are $4~\%$ and $8~\%$ larger than those observed from the coarse- and intermediate-resolution simulations, respectively.

\begin{figure}[htbp!] 
\centering \includegraphics[width=\textwidth]{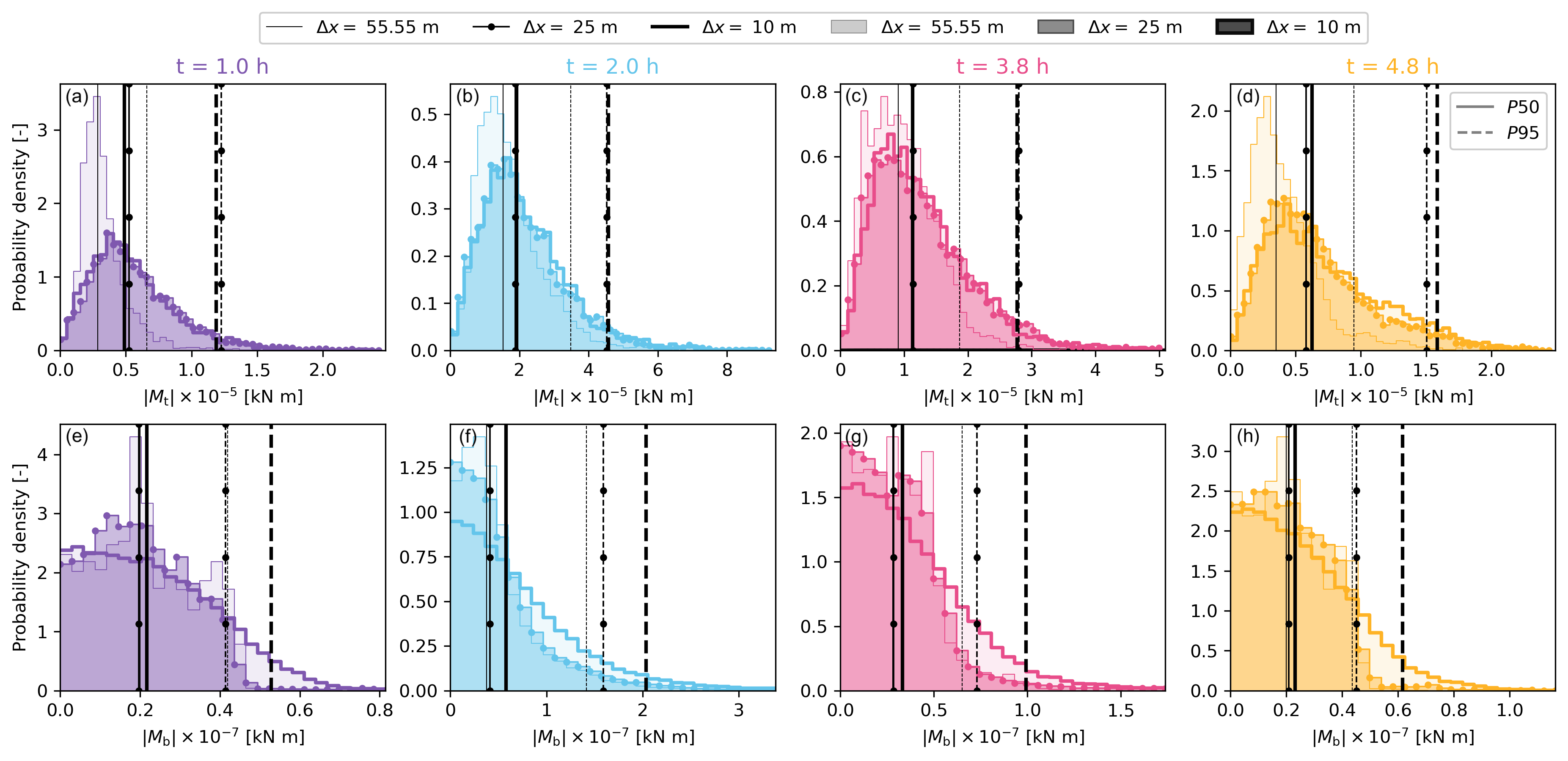}
\caption{Probability density distributions of the tower-base (a--d) and blade-root moments (e--h) from the coarse- and fine-resolution inflow conditions for 10 min periods centered at $t\approx1\,{\rm h}$ (a,e), $t\approx2\,{\rm h}$ (b,f), $t=3.8\,{\rm h}$ (c,g), and $t=4.8\,{\rm h}$ (d,h). The vertical solid and dashed lines in each panel represent the $50^{\rm th}$ and $95^{\rm th}$ percentiles, respectively, for each load distribution. }
\label{fig:loads_quantiles}
\end{figure}

To further compare the loads resulting from each inflow condition, we examine the load distribution. The distributions of the tower-base and blade-root bending moments are evaluated for four 10 min periods centered at $t\approx1\,{\rm h}$, $t\approx2\,{\rm h}$, $t=3.8\,{\rm h}$, and $t=4.8\,{\rm h}$. Figure \ref{fig:loads_quantiles} shows the probability density of the tower-base (Fig. \ref{fig:loads_quantiles}a--d) and blade-root (Fig. \ref{fig:loads_quantiles}e--h) bending moments resulting from the coarse- (light shades, thin lines), intermediate- (semi-dark shades, thin lines with circled markers), and fine-resolution (dark shades, thick lines) inflow conditions. Furthermore, we estimate the $50^{\rm th}$ and $95^{\rm th}$ percentiles for each distribution. The $50^{\rm th}$ percentile (i.e., median) provides the typical loading throughout each 10 min period while the $95^{\rm th}$ percentile illustrates the most extreme load events. In Fig. \ref{fig:loads_quantiles}, the vertical solid and dashed lines indicate the $50^{\rm th}$ $(P50)$ and $95^{\rm th}$ $(P95)$ percentiles, respectively. We also evaluate the difference in the $50^{\rm th}$ and $95^{\rm th}$ percentiles between the tower-base moments resulting from each inflow condition using a nonparametric bootstrap resampling method to characterize uncertainty. We conclude that the percentiles are statistically distinct when the $95~\%$ confidence intervals associated with each inflow condition do not overlap.

The tower-base and blade-root load distributions from the fine-resolution inflow exhibit consistently larger median ($50^{\rm th}$ percentile) and extreme ($95^{\rm th}$ percentile) loads than those from the coarse-resolution inflow throughout the hurricane passage (Fig. \ref{fig:loads_quantiles}). The load distributions from both the coarse and fine LES inflow are right-skewed; however, the distributions from the fine-resolution inflow exhibit longer right tails due to more extreme loads. Differences in the tower-base and blade-root load distributions are larger far from the storm's eyewall (Fig. \ref{fig:loads_quantiles}a,d) than near the eyewall (Fig. \ref{fig:loads_quantiles}c,d). The $50^{\rm th}$ and $95^{\rm th}$ percentiles at $t\approx1\,{\rm h}$ and $t=4.8\,{\rm h}$ in the tower-base loads from the coarse-resolution inflow can be about $40~\%$ smaller than those resulting from the fine-resolution inflow conditions, whereas the median and extreme blade-root loads are about $15~\%$ and $25~\%$ smaller, respectively. Close to the eyewall (at $t\approx2\,{\rm h}$ and $t=3.8\,{\rm h}$), the $50^{\rm th}$ percentile in the tower-base (blade-root) load distribution from the coarse-resolution inflow is about $20~\%$ (between $15~\%$ and $35~\%$) smaller than those from the fine-resolution inflow. The $95^{\rm th}$ percentile near the eyewall can be between 20~\% and 30~\% smaller when using the coarse-resolution inflow for both the tower-base and blade-root moments. In all cases, differences in both $50^{\rm th}$ and $95^{\rm th}$ percentiles are statistically significant. This means the coarse LES inflow is inadequate to capture the typical and extreme tower-base and blade-root loading throughout the hurricane passage.

Although the blade-root load distributions from the fine LES inflow exhibit larger median and extreme loads than those from the intermediate LES inflow, the tower-base load distributions can be well represented by the intermediate LES inflow, especially near the eyewall (Fig. \ref{fig:loads_quantiles}). Like with the coarse LES inflow, differences in the load distributions are also smaller near the eyewall. Differences in the $50^{\rm th}$ and $95^{\rm th}$ percentiles at $t\approx1\,{\rm h}$ and $t=4.8\,{\rm h}$ in the tower-base loads from the intermediate-resolution inflow are between $3~\%$ and $7~\%$ compared to those resulting from the fine-resolution inflow conditions. Conversely, differences in the median and extreme loads for the blade-root moments at the same times are $10~\%$ and $25~\%$, respectively. Close to the eyewall, the $50^{\rm th}$ and $95^{\rm th}$ percentiles of the tower-base moments from the intermediate and fine LES are within $1~\%$ of each other. Differences in the median and extreme blade-root moments near the eyewall are between $20~\%$ and $30~\%$. Therefore, the intermediate-resolution inflow can yield accurate estimates of tower-base loading during the eyewall hit but the blade-root load distribution are not well captured throughout the hurricane passage.

\begin{figure}[htbp!] 
\centering \includegraphics[width=\textwidth]{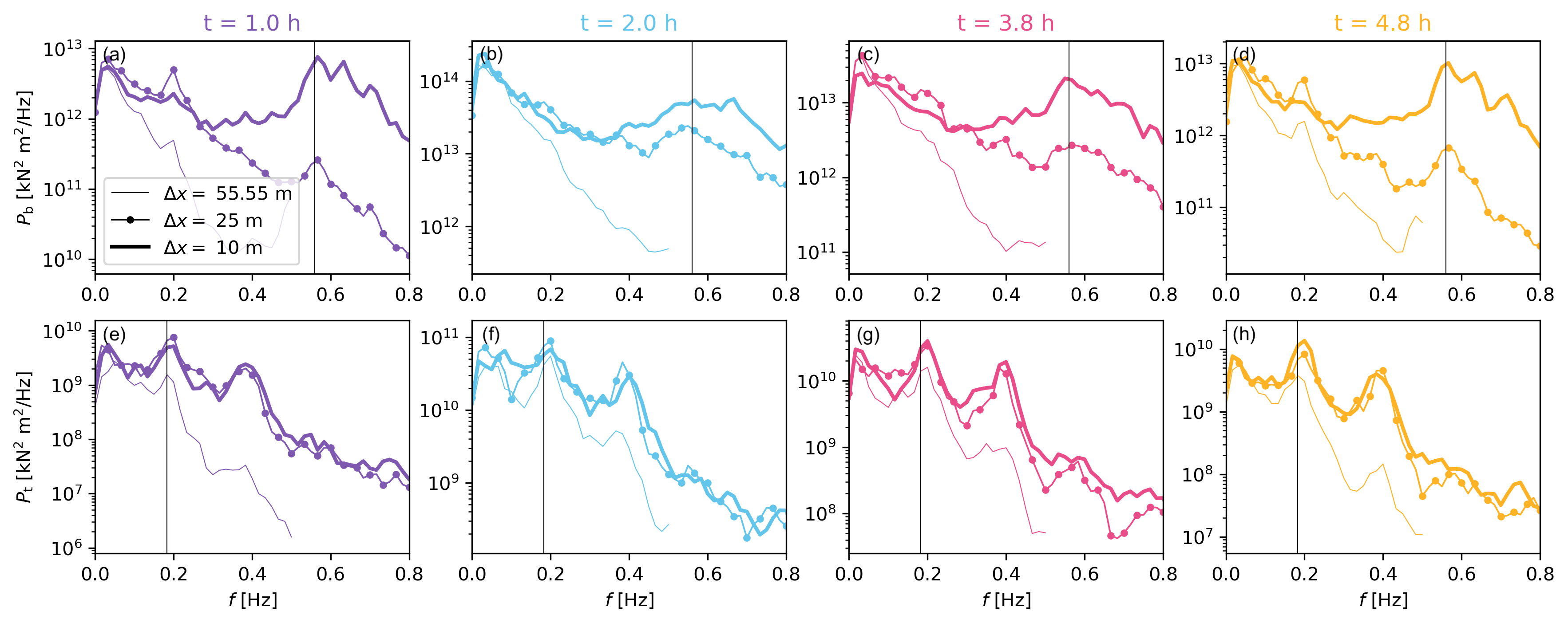}
\caption{Frequency response of the blade root (a--d) and tower base (e--h) moments. Panels (a,e) correspond to $t\approx1\,{\rm h}$ (purple), panels (b,f) correspond to $t\approx2\,{\rm h}$ (light blue), panels (c,g) correspond to $t=3.8\,{\rm h}$ (pink), and panels (d,h) correspond to $t=4.8\,{\rm h}$ (yellow). The vertical black line in panels (a--d) and (e--h) represent the natural frequency of the blades $(f_{n,b}=0.56\,{\rm Hz})$ and tower $(f_{n,t}=0.17\,{\rm Hz})$, respectively.}
\label{fig:loads_turb_freq}
\end{figure}

To understand the limitations of the coarse- and intermediate-resolution inflow conditions in eliciting the correct turbine response, we examine the frequency response of the loads at the four times of interest (colored lines in Fig. \ref{fig:loads_turb_freq}). The power spectral density of the tower-base $(P_t)$ and blade-root  $(P_b)$ moments are estimated over a 10 min window, centered at the four times of interest shown in Fig. \ref{fig:loads_turb_time}a. To quantify differences in the loads in the frequency space, we estimate the variance contained within a frequency band defined by $f_1$ and $f_2$ as $\sigma^2_{i}=\int_{f_1}^{f_2} P_i\,df$. Because the coarse-resolution inflow is at $1\,{\rm Hz}$, we only include turbine load data from the coarse LES inflow for frequencies smaller than the Nyquist frequency (i.e., $0.5\,{\rm Hz}$).

The frequency response of the turbine loads changes with the inflow conditions (Fig. \ref{fig:loads_turb_freq}). In general, the loads from the aeroelastic simulations with higher-resolution inflow contain larger variance at high frequencies. Because the coarse-resolution inflow cannot adequately resolve small turbulence scales (i.e., frequencies larger than $\sim0.1$--$0.2\,{\rm Hz}$; see Sect \ref{sec:turbSpctr}), the variance of the tower-base and blade-root moments obtained from the coarse-resolution inflow are highly underestimated for frequencies $f\gtrapprox0.1\,{\rm Hz}$. The variance in the blade-root moments $(\sigma^2_b)$ contained within $(f_1,f_2)=(0.1\,{\rm Hz},0.5\,{\rm Hz})$ is $82~\%$, $68~\%$, $71~\%$, and $76~\%$ larger for the fine LES inflow compared to the coarse LES inflow for the four times of interest, respectively. Similarly, the variance of the tower-base moment $(\sigma^2_t)$ contained within $(f_1,f_2)=(0.1\,{\rm Hz},0.5\,{\rm Hz})$ is $80~\%$, $58~\%$, $67~\%$, and $75~\%$ larger for the fine LES inflow compared to the coarse LES inflow for the four times of interest, respectively. In a similar vein, because the intermediate-resolution inflow cannot resolve turbulence scales with frequencies larger than $\sim0.2$--$0.3\,{\rm Hz}$ (see Sect \ref{sec:turbSpctr}), the load variance obtained from the intermediate-resolution inflow can be underestimated for frequencies $f\gtrapprox0.3\,{\rm Hz}$, especially for the blade-root moments. The variance in the blade-root moments $(\sigma^2_b)$ contained within $(f_1,f_2)=(0.3\,{\rm Hz},0.8\,{\rm Hz})$ is $93~\%$, $58~\%$, $77~\%$, and $91~\%$ larger for the fine LES inflow compared to the intermediate LES inflow for the four times of interest, respectively. The variance of the tower-base moment $(\sigma^2_t)$ contained within $(f_1,f_2)=(0.3\,{\rm Hz},0.8\,{\rm Hz})$ is $16~\%$, $12~\%$, $37~\%$, and $10~\%$ larger for the fine LES inflow compared to the intermediate LES inflow for the four times of interest, respectively. Thus, because the coarser-resolution simulations may not resolve small-scale turbulence, the inflow wind fields cannot excite those frequencies in the turbine response.

More important than the load variance contained within broad frequency bands is the load variance contained within narrow bands centered on the structure's natural frequencies. Here, we consider the load variance contained around the natural frequency of the tower $(f_{n,t} = 0.17\,{\rm Hz})$ and the blades $(f_{n,t}=0.56\,{\rm Hz})$. The natural frequencies of the tower and blades are shown as vertical black lines in Fig. \ref{fig:loads_turb_freq}. We use the tower-base and blade-root load variance contained within $(f_1,f_2)=(0.16\,{\rm Hz},0.2\,{\rm Hz})$ and $(f_1,f_2)=(0.53\,{\rm Hz},0.58\,{\rm Hz})$, respectively, in the simulation with fine LES inflow as a baseline for comparison.

The coarse LES inflow cannot excite the natural frequencies of the tower and blades, underestimating the associated load magnitude on the turbine structure (Fig. \ref{fig:loads_turb_freq}). The variance of the tower-base moment within $(f_1,f_2)=(0.16\,{\rm Hz},0.2\,{\rm Hz})$ is $73~\%$, $29~\%$, $57~\%$, and $70~\%$ larger in the simulation with fine-resolution inflow for the four times of interest, respectively (Fig. \ref{fig:loads_turb_freq}e--h). The variance of the blade-root moment within $(f_1,f_2)=(0.53\,{\rm Hz},0.58\,{\rm Hz})$ in the simulation with coarse LES inflow is negligible compared to that of the fine-resolution inflow for the four times of interest (Fig. \ref{fig:loads_turb_freq}e--h). Thus, because the coarse LES inflow contains minimal energy in the flow around $f_{n,t} = 0.17\,{\rm Hz}$ and virtually no energy at $f_{n,b} = 0.56\,{\rm Hz}$ (Fig. \ref{fig:spectra_ui}), it cannot excite the structure's natural frequencies, leading to underestimation of the turbine loads throughout the hurricane passage.

The intermediate-resolution inflow can excite the natural frequencies of the tower but not those of the blades (Fig. \ref{fig:loads_turb_freq}). The intermediate LES accurately excites the tower's natural frequency, where differences in the load variance contained between $(f_1,f_2)=(0.16\,{\rm Hz},0.2\,{\rm Hz})$ are $4~\%$, on average, across all times (Fig. \ref{fig:loads_turb_freq}a--d). Conversely, the load variance contained near the blades' natural frequency is highly underestimated. The variance of the blade-root moment within $(f_1,f_2)=(0.53\,{\rm Hz},0.58\,{\rm Hz})$ in the simulation with the intermediate LES inflow is $96~\%$, $57~\%$, $86~\%$, and $93~\%$ larger in the simulation with fine-resolution inflow for the four times of interest (Fig. \ref{fig:loads_turb_freq}e--h). To summarize, the intermediate LES inflow contains sufficient energy at small turbulent scales to excite the tower's natural frequency $f_{n,t} = 0.17\,{\rm Hz}$ but not for the blades ($f_{n,b} = 0.56\,{\rm Hz}$) (Fig. \ref{fig:spectra_ui}); therefore, the load response of the tower can be well modeled, whereas the loads on the blades may be underestimated.

\section{Discussion and conclusions} \label{sec:concl}  
In this work, we performed long-duration turbulence-resolving simulations of a hurricane before it makes landfall and evaluated a wind turbine's response to the passage of the storm. We conducted LES of Hurricane Laura (2020) using a coarse $(\Delta x=55.55\,{\rm m})$, an intermediate $(\Delta x=25\,{\rm m})$, and a fine $(\Delta x=10\,{\rm m})$ grid to study differences in the mean wind and turbulence structures that develop as the hurricane moves over a virtual wind turbine position. Then, we simulated the aeroelastic response of a wind turbine to the hurricane passage, using direct inflow from the LESs. We found critical differences in the flow structure that develops within each grid, as well as the wind turbine's response from each inflow condition.

The passage of Hurricane Laura causes extreme wind conditions at turbine heights that can produce extreme loads on a wind turbine's structure. As the hurricane approaches the virtual turbine position, the streamwise wind speed (mean wind speed and gusts) and the velocity variances increase in magnitude. Even though the wind speed and turbulence change dramatically, the overall vertical structure of the wind profile changes minimally as the leading edge of the eyewall approaches the turbine. Wind speed shear remains nearly unchanged, but wind veer tends to increase. Similarly, changes in the size of the dominant eddies in the flow across the turbine rotor layer are small over time. Nevertheless, the dramatic increase in wind speed and turbulence content of the flow around the leading edge of the hurricane eyewall produces extreme loads on the tower base and blade root of a wind turbine. Wind speed and turbulence suddenly decrease in the eye of the storm, also reducing the maximum loads on the turbine structure. As the trailing edge of the storm's eyewall moves over the turbine position, wind speed and turbulence content increase at the same time as wind speed shear and wind veer. Nonetheless, the size of the dominant eddies in the flow remains nearly unchanged. The increase in wind speed and turbulence content of the flow also increases the mean and maximum loads on the tower base and blade root of the turbine; however, the loads are smaller than at the leading edge of the storm's eyewall.

Mean wind, wind gusts, and turbulence statistics at turbine-relevant heights can vary significantly between the coarse-, intermediate, and fine-resolution LESs of Hurricane Laura. Both the mean wind and turbulence statistics differ between the coarse- and fine-resolution wind fields; however, the hub-height wind speed remains nearly the same. The mean wind and turbulence statistics remain comparable between the intermediate- and fine-resolution LESs. Vertical shear of wind speed and wind direction of the mean wind and gusts are typically larger in the coarse-resolution simulation compared to the intermediate and fine LES, generally driven by increased shear below hub height. Furthermore, the coarse-resolution grid generally underestimates the magnitude of the velocity variances as well as the distribution of energy across scales. The coarse-resolution LES overestimates the variance in the flow contained within low frequencies, likely due to reduced energy transfer across scales \citep{sanchezgomez_turbulenceresolving_2025}. As a result, the dominant eddies in the flow are consistently larger in the coarse-resolution simulation. The coarse-resolution simulation also does not adequately capture the inertial subrange of the flow, especially for the vertical velocity component. Our simulation results are similar to those of Sanchez-Gomez et al. \cite{sanchezgomez_turbulenceresolving_2025}; however, the differences in turbulence structures are larger here because the coarse-resolution LES employs a larger horizontal grid spacing. The intermediate-resolution LES generally captures the magnitude of the velocity variances and better represents the size of the dominant turbulent eddies in the flow. However, the grid resolution in the intermediate LES is not fine enough to accurately capture the inertial subrange of the flow at hub height throughout the entire hurricane passage. Collectively, our results show that the coarse grid is incapable of accurately capturing the mean wind and turbulence structure of Laura during a direct eyewall hit at turbine-relevant heights, whereas the intermediate grid captures the bulk flow behavior while underestimating the energy for small-scale turbulence. 

Differences in mean wind and turbulence structures between microscale flow fields of varying fidelity produce critical differences in the wind turbine loads experienced during a hurricane passage. The combination of smaller coherent eddies, larger variances contained within the small-scale turbulence, and larger variances overall in the turbine rotor layer can generate extreme loads on a wind turbine. Notably, since the coarse-resolution inflow contains minimal energy at the tower’s and blades' natural frequencies, it cannot effectively excite those structural modes. The intermediate-resolution inflow can excite the natural frequencies of the tower and can yield adequate estimates of the load magnitude; however, the intermediate-resolution inflow contains minimal energy at the blades' natural frequency, resulting in load underestimations. Kapoor et al. \cite{kapoor_hurricane_2020} showed that wind veer can amplify loads on idling wind turbines during hurricane conditions. Surprisingly, the coarse-resolution inflow, which has larger wind speed and direction shear across the turbine rotor layer, does not cause larger mean and maximum loads on the turbine structure. This suggests the energy contained within the smaller scales of turbulence, their correlation across space and time, and the distinct mean profiles that they produce are crucial to idling wind turbine load estimations. 

One critical finding of our study is that the grid resolution required to generate wind fields for idling wind turbine loads analyses is a function of the structure's natural frequencies. The wind fields must resolve the small-scale turbulent structures that excite the natural frequencies of the turbine tower and blades. For a structure with natural frequency $f_{n,i}$, a flow field with characteristic hub-height wind speed $U_h$, and a numerical model with effective grid resolution $\Delta x_{eff}\approx\lambda \Delta x$, the largest grid spacing of the numerical model can be approximated by $\Delta x_{\rm max} \approx \frac{U_h}{\lambda f_{n,i}}$. The effective resolution of the numerical models considered here is $\Delta x_{eff}=\lambda \Delta x\approx 7\Delta x$. Conducting idling wind turbine loads analyses employing wind fields that are not sufficiently finely resolved can lead to the underestimation of the extreme load magnitudes (Fig. \ref{fig:loads_quantiles}) and the frequency response of the structure (Fig. \ref{fig:loads_turb_freq}). Thus, even if the mean wind and turbulence content of the flow are well captured by given simulation (like our $\Delta x=25\,{\rm m}$ case), the wind turbine's response may not be well represented. Our results also indicate that $\Delta x=10\,{\rm m}$ may not be sufficient to study the wind turbine's response to the passage of weaker storms or for locations farther away from the eyewall, where the characteristic wind speed of the flow $U_h$ is slower.

This study has provided the most detailed analysis to date on the mean wind structure and turbulence evolution of a hurricane at turbine-relevant heights. It is also the first analysis on a wind turbine's response to a realistic hurricane passage. Nonetheless, our analysis is limited to wind conditions and the turbine's response to a direct eyewall hit. It is unlikely that a tropical cyclone would move exactly over a wind turbine position. The wind turbine's response to the passage of a tropical cyclone at different distances from the storm's center and track will be presented in Part II of this study. We will also study differences in structural loads between the hurricane passage and the IEC wind turbine design conditions in Part II. Furthermore, we only consider one tropical cyclone in our analysis. Because different storms of the same intensity can have different structures and because the intensity of the storm will also change the turbine loads, a more thorough analysis covering a wider range of storms is advised. As storms of varying intensities are analyzed, the sensitivity of the wind fields to the grid spacing of the simulations should also be analyzed. As shown here, a coarser grid may be sufficient to adequately capture wind conditions at the eyewall of a high-intensity tropical cyclone; however, less intense storms may require smaller grid spacing. These high-resolution simulations of the hurricane passage can also be used to derive wind turbine control strategies to mitigate wind turbine loads during a hurricane passage and evaluate the risk that hurricanes pose on other components of the energy infrastructure, such as transmission lines, photovoltaic systems, and oil rigs as well as civil infrastructure. 

\section*{Data and code availability}
The OpenFAST code used in this study is publicly available at \url{https://github.com/OpenFAST/openfast}. The FastEddy\textsuperscript{\textregistered} code is publicly available at \url{https://github.com/NCAR/FastEddy-model}. The modified WRF source code is available at \url{https://github.com/miguel-sg-2/WRF_versions}.

The microscale simulation data at the different grid spacings is publicly available at \url{https://doi.org/10.21947/3367311}. 
\section*{Acknowledgements}
This work was authored by the National Laboratory of the Rockies for the U.S. Department of Energy (DOE), operated under Contract No. DE-AC36-08GO28308. Funding provided by U.S. Department of Energy Office of Critical Minerals and Energy Innovation Integrated Energy Systems Office. The views expressed in the article do not necessarily represent the views of the DOE or the U.S. Government. The U.S. Government retains and the publisher, by accepting the article for publication, acknowledges that the U.S. Government retains a nonexclusive, paid-up, irrevocable, worldwide license to publish or reproduce the published form of this work, or allow others to do so, for U.S. Government purposes. This research was performed using computational resources sponsored by the U.S. Department of Energy's Office of Critical Minerals and Energy Innovation and located at the National Laboratory of the Rockies.

\appendix
\section{Convergence of wind fields and loads \label{sec:appA}}   
To examine sensitivity to the grid spacing of the fine-resolution LES, we conduct additional simulations at a finer grid spacing. LES at $\Delta x=5 \,{\rm m}$ are conducted to evaluate sensitivity to the wind field structure and the turbine's response. Due to the high computational cost and data storage requirements of these finer-resolution simulations, we only run the LES for half of the storm passage (i.e., 2 h, 45 min). Table \ref{app:tab:numConfig} summarizes the domain configuration of the $\Delta x=5 \,{\rm m}$ microscale simulations. In the forthcoming analysis, we first compare the mean wind and turbulence characteristics of the flow between the $\Delta x=5 \,{\rm m}$ and $\Delta x=10 \,{\rm m}$ LESs. Then, we consider the wind turbine's response associated with each inflow condition.

\begin{table*}[t]
\caption{Numerical domain configuration of the finest-resolution microscale simulation, including the number of grid points along each $i$-direction $(n_i)$, domain size $(L_i)$, horizontal grid spacing $(\Delta x)$, mean vertical grid spacing across the turbine rotor layer $(\Delta z_{\rm RL})$, and time step $(\Delta t)$. } %
\resizebox{\textwidth}{!}{
\begin{tabular}{lcccccc}
\hline
\hline
Numerical Model & Domain & $(n_x,n_y,n_z)$ $[-]$ & $(L_x,L_y,L_z)$ $[\rm km]$ & $\Delta x$ $[\rm m]$ & $\langle \Delta z \rangle_{\rm RL}$ $[\rm m]$  & $\Delta t$ $[\rm s]$ \\
\hline
FastEddy & Finest & $(3000,2970,122)$ & $(15,14.85,1)$ & $5$ & $3.5$  & $0.005$ \\ %
\hline
\hline
\end{tabular}
}
\label{app:tab:numConfig}
\end{table*}

\subsection{Wind fields} 
\begin{figure}[htbp!] 
\centering \includegraphics[height=8cm]{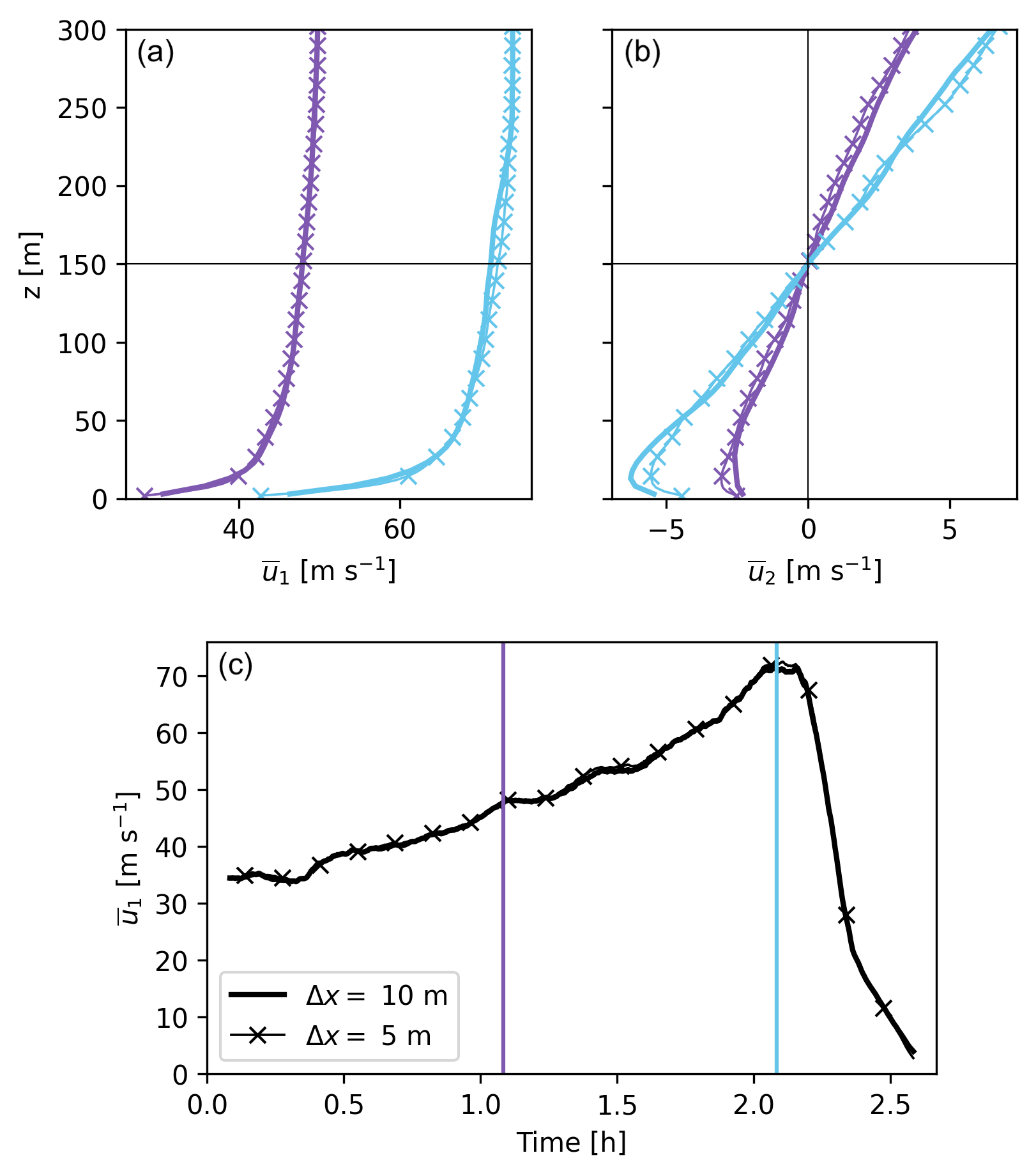}
\caption{Mean streamwise (a) and cross-stream (b) wind profiles. Panel (c) shows the temporal evolution of the mean horizontal wind speed at $z=150\,{\rm m}$ at the virtual turbine position. The vertical profiles in panels (a) and (b) are sampled at the two times shown in panel (c). }
\label{app:fig:meanWinds}
\end{figure}

To evaluate convergence of the wind field structure, we consider the 10 min averaged wind profiles, the velocity variances, and the energy spectra from the $\Delta x=10\,{\rm m}$ and $\Delta x=5\,{\rm m}$ simulations. Furthermore, because the $\Delta x=5\,{\rm m}$ simulations are run for half of the storm passage, we focus on two of the four representative times described in Section \ref{sec:hurr_winds}. Specifically, we analyze the wind structure when the hurricane is far south of the virtual turbine ($t\approx1\,{\rm h}$; purple line in Fig. \ref{app:fig:meanWinds}), and when the leading edge of the eyewall reaches the turbine position ($t\approx2\,{\rm h}$; light-blue line in Fig. \ref{app:fig:meanWinds}).

The mean wind structure is virtually unchanged between the $\Delta x=10\,{\rm m}$ and $\Delta x=5\,{\rm m}$ simulations (Fig. \ref{app:fig:meanWinds}). Differences in the mean hub-height streamwise wind speed between both simulations are always less than 2.5~\% and are on average less than 1~\% throughout the simulation time (Fig. \ref{app:fig:meanWinds}c). Likewise, the vertical wind structure remains comparable between both simulations. The power-law exponent fit to the simulated wind profiles is on average $\alpha=0.053$ and $\alpha=0.061$ for the $\Delta x=10\,{\rm m}$ and $\Delta x=5\,{\rm m}$ simulations, respectively. Differences in vertical wind direction shear are even smaller: mean wind veer across the rotor layer is $7^\circ$ in the $\Delta x=10\,{\rm m}$ LES, and it is $7.08^\circ$ for the $\Delta x=5\,{\rm m}$ LES.

\begin{figure}[htbp!] 
\centering \includegraphics[height=10cm]{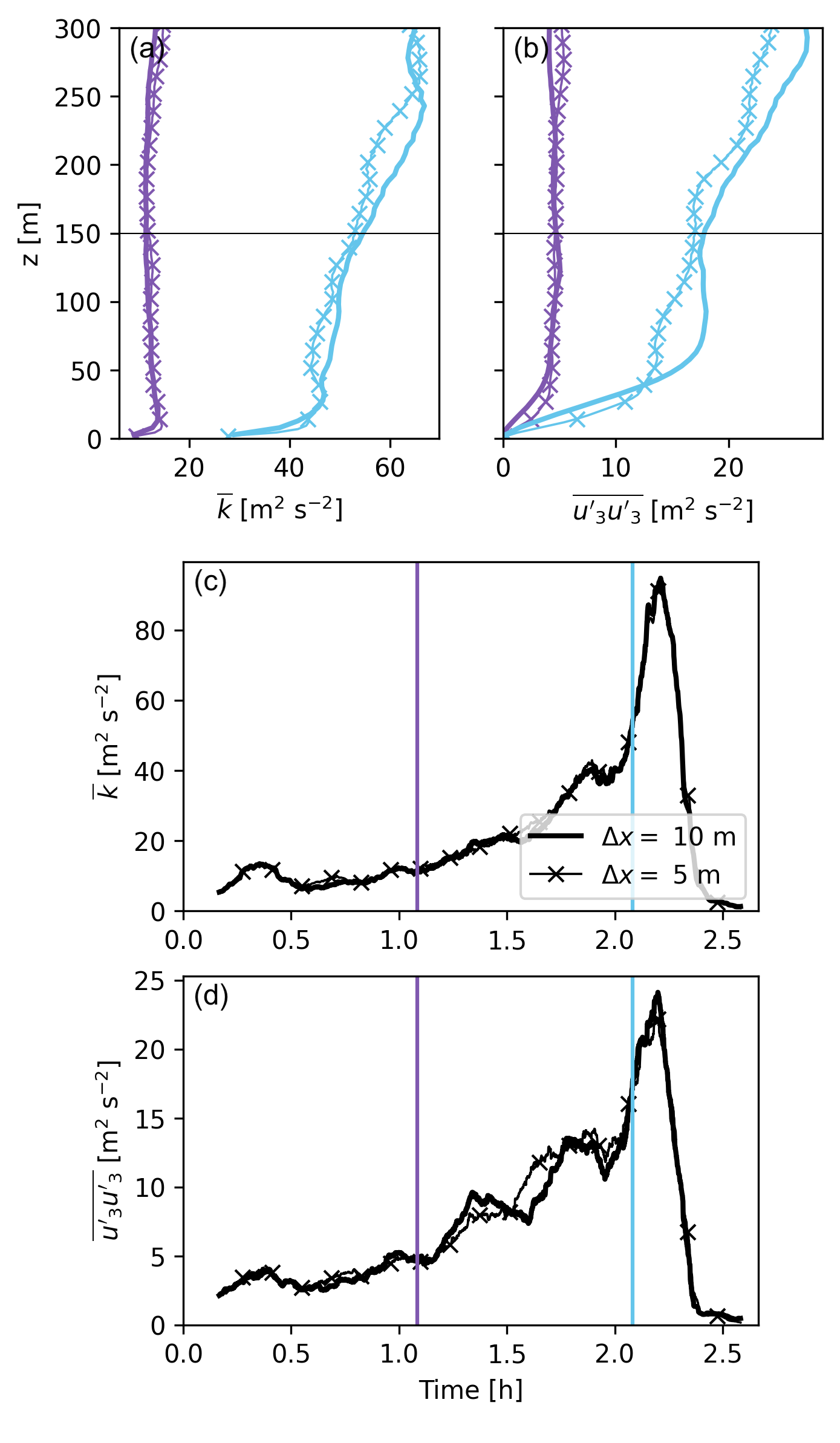}
\caption{Vertical profiles of turbulence kinetic energy (a) and vertical velocity variance (b) in the lowest $300\,{\rm m}$ of the boundary layer. The temporal evolution of the turbulence kinetic energy and vertical velocity variance at $z=150\,{\rm m}$ at the virtual turbine position are shown in panels (c) and (d), respectively. The vertical profiles in panels (a,b) are sampled at the two times shown in panels (c,d). }
\label{app:fig:turb}
\end{figure}

Turbulence content of the flow also remains similar between the $\Delta x=10\,{\rm m}$ and $\Delta x=5\,{\rm m}$ simulations (Fig. \ref{app:fig:turb}). To represent the turbulence content of the flow in all three directions, we estimate the turbulence kinetic energy, defined as half the trace of the Reynolds stress tensor $(\overline{k}=\frac{1}{2}\overline{u'_iu'_i})$. On average throughout the simulation time, the resolved hub-height turbulence kinetic energy in the flow and the vertical velocity variance in both simulations differ by 5.5~\% and 8.5~\%, respectively. Note that differences in the resolved velocity variances are expected due to contributions from the subgrid-scale turbulence model. Nevertheless, the turbulent kinetic energy and vertical velocity variance profiles display comparable magnitudes and vertical distributions. The $\Delta x=5\,{\rm m}$ simulation also confirms that the size of the dominant eddies in the flow is being well captured by the $\Delta x=10\,{\rm m}$ LES. The integral length scales of the flow at hub height for the streamwise, cross-stream, and vertical wind components along the cross-stream direction are respectively $L_{u_1}^{y^*}=135\,{\rm m}$, $L_{u_2}^{y^*}=157\,{\rm m}$, and $L_{u_3}^{y^*}=76\,{\rm m}$, on average, for the $\Delta x=5\,{\rm m}$ simulation. Similarly, the integral length scales of the flow for the streamwise, cross-stream, and vertical wind components along the cross-stream direction are respectively $L_{u_1}^{y^*}=126\,{\rm m}$, $L_{u_2}^{y^*}=163\,{\rm m}$, and $L_{u_3}^{y^*}=68\,{\rm m}$, on average, for the $\Delta x=10\,{\rm m}$ simulation.

\begin{figure}[htbp!] 
\centering \includegraphics[height=7cm]{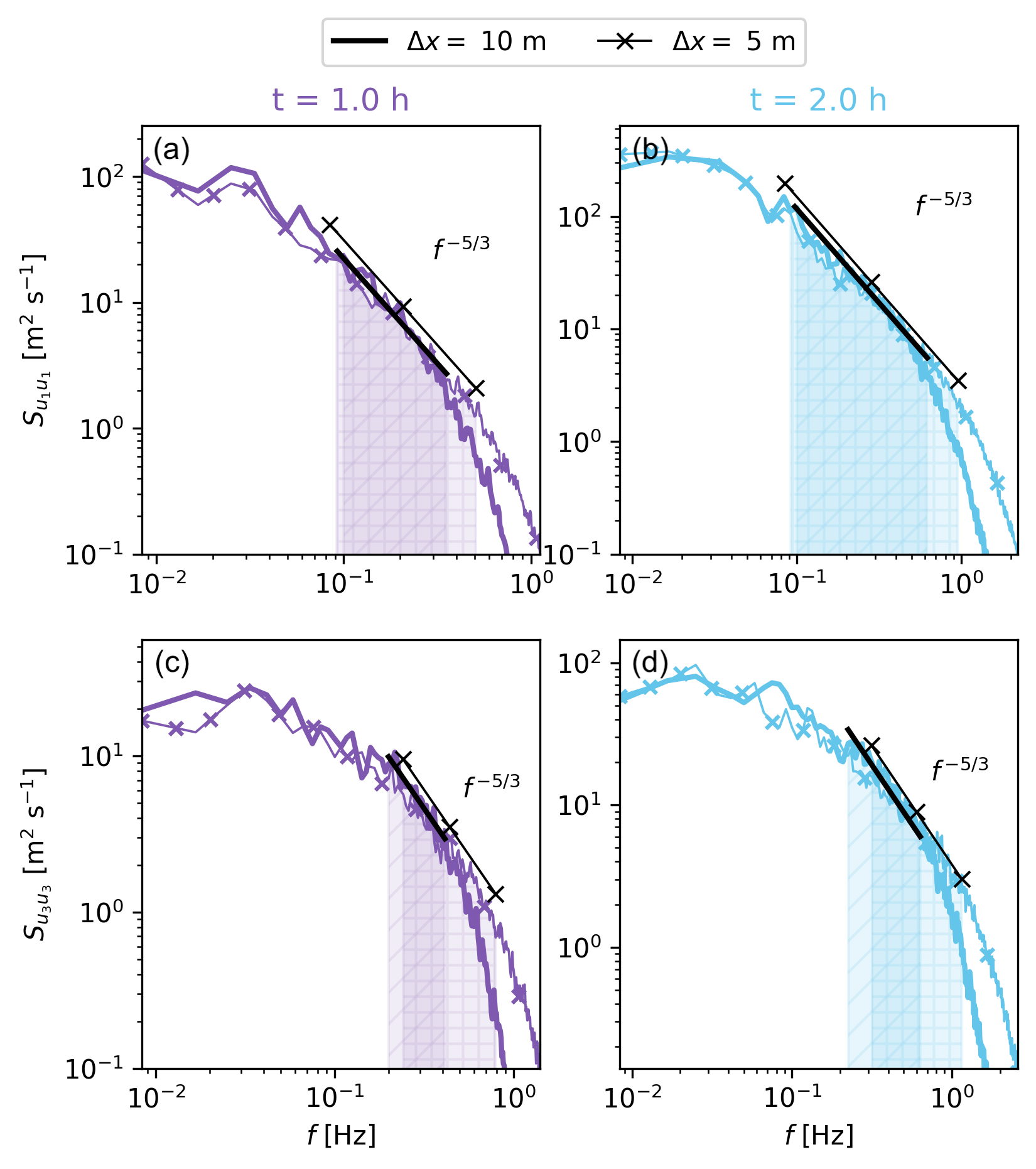}
\caption{Energy spectra of the streamwise (a--d), and vertical (e--h) wind components. Results are shown for the two times of interest shown in Fig. \ref{app:fig:meanWinds}. The hatched areas under each line represent the range of frequencies that characterize the inertial subrange (i.e, $S_{ii}\sim f^{-5/3}$) for each grid.}
\label{app:fig:spctr}
\end{figure}

The $\Delta x=10\,{\rm m}$ simulation adequately captures the energy distribution across scales before and during the eyewall hit (Fig. \ref{app:fig:spctr}). As expected, the finer grid captures higher frequencies in the flow. However, the energy spectra of the streamwise and vertical wind components remain nearly unchanged for frequencies $f\lessapprox0.4\,{\rm Hz}$ before the eyewall hit and for frequencies $f\lessapprox0.6\,{\rm Hz}$ as the eyewall impacts the turbine position. This indicates that the energy transfer across scales is well captured in the $\Delta x=10\,{\rm m}$ simulation and that the finer grid resolution is simply extending the resolvable scales within the inertial subrange without significantly modifying the wind field structure.

\subsection{Wind turbine loads}   
Having shown that the wind field structure remains virtually unchanged between the $\Delta x=10\,{\rm m}$ and $\Delta x=5\,{\rm m}$ simulations, we now study the wind turbine loads associated with each inflow condition. As explained in Sect. \ref{sec:aeroelast_sims}, we sample the velocity fields on a volume that fully encompasses the turbine. The three-dimensional velocity fields of the $\Delta x=5\,{\rm m}$ are saved at $8\,{\rm Hz}$ to match the smaller grid spacing.

\begin{figure}[htbp!] 
\centering \includegraphics[width=0.5\textwidth]{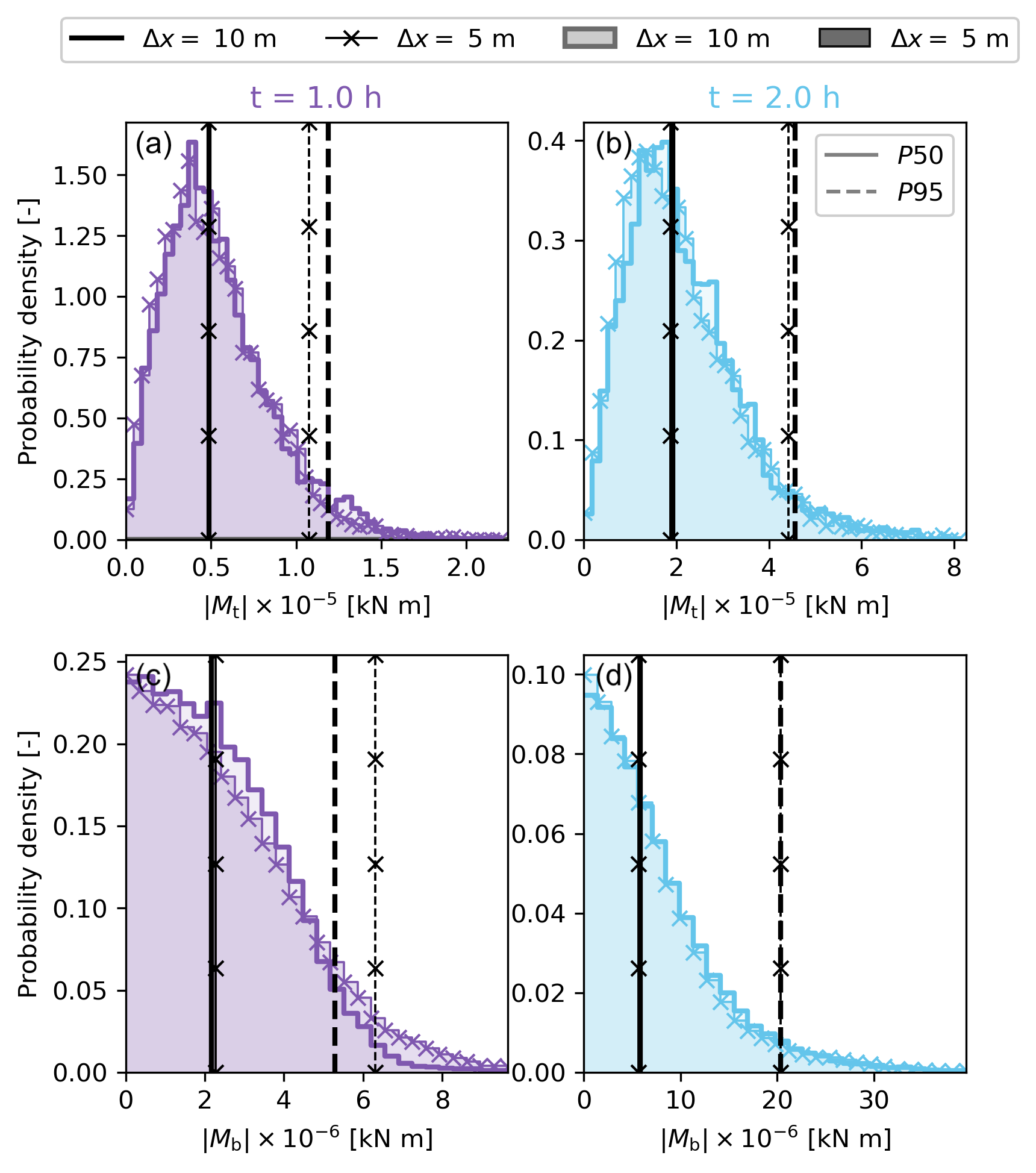}
\caption{Probability density distributions of the tower-base moment (a,b) and the blade-root moments (c,d) from the fine- and finest-resolution inflow conditions for 10 min periods centered at $t\approx1\,{\rm h}$ (a,c), and $t\approx2\,{\rm h}$ (b,d). The vertical solid and dashed lines in each panel represent the $50^{\rm th}$ and $95^{\rm th}$ percentiles, respectively, for each load distribution. }
\label{app:fig:loads_quantiles}
\end{figure}

The load distributions for the tower-base and blade-root bending moments are similar for both inflow conditions, especially as the hurricane eyewall impacts the turbine (Fig. \ref{app:fig:loads_quantiles}). The load distributions for the tower-base and blade-root bending moments are statistically equal when the leading eyewall of the hurricane impacts the turbine position (Fig. \ref{app:fig:loads_quantiles}b,d). The $50^{\rm th}$ and $95^{\rm th}$ percentiles for both the tower-base and blade-root moments are within 3~\% of each other during the eyewall hit (Fig. \ref{app:fig:loads_quantiles}b,d). When the hurricane is far south of the turbine, the load distribution of the tower-base bending moment remains well captured for both inflow conditions (Fig. \ref{app:fig:loads_quantiles}a). Although the $50^{\rm th}$ percentile is 15~\% larger in the $\Delta x=5\,{\rm m}$ simulation, the $95^{\rm th}$ percentiles are within 3~\% of each other (Fig. \ref{app:fig:loads_quantiles}a). However, the load distribution for the blade-root bending moment shows important differences, especially for the extreme values (Fig. \ref{app:fig:loads_quantiles}c). The $50^{\rm th}$ percentile for the blade-root bending moments are within 2~\% of each other, while the $95^{\rm th}$ percentiles differ by about 18~\% (Fig. \ref{app:fig:loads_quantiles}c).

\begin{figure}[htbp!] 
\centering \includegraphics[height=7cm]{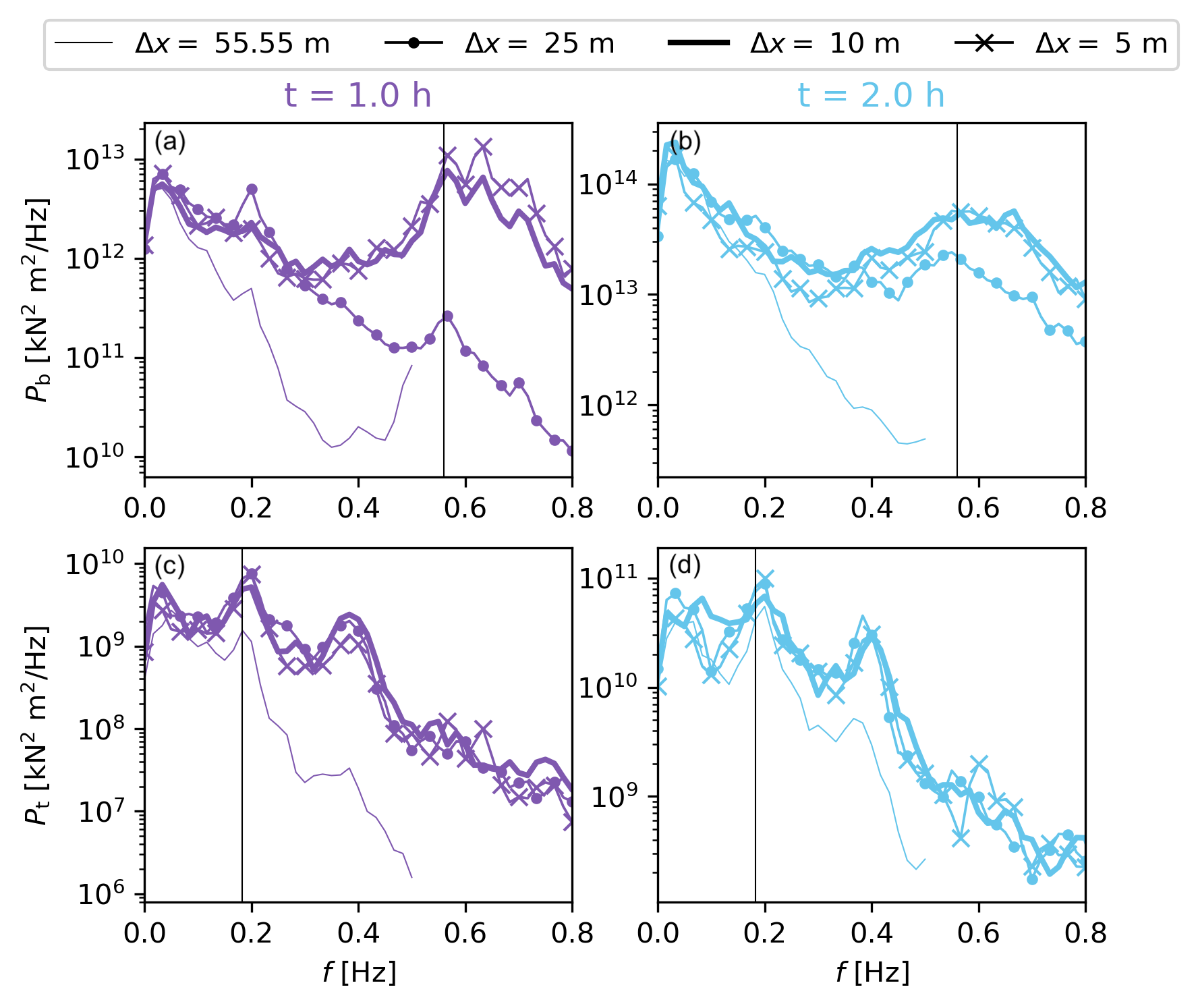}
\caption{Frequency response of the blade root (a,b) and tower base (c,d) moments. Panels (a,b) correspond to $t\approx1\,{\rm h}$, and panels (c,d) correspond to $t\approx2\,{\rm h}$. The vertical black line in panels (a,b and (c,d) represents the resonant frequency of the blades $(f_r=0.56\,{\rm Hz})$ and tower $(f_r=0.17\,{\rm Hz})$, respectively. Results for the $\Delta x=55.55\,{\rm m}$ and $\Delta x=25\,{\rm m}$ simulations are also shown for reference.}
\label{app:fig:loads_turb_freq}
\end{figure}

The frequency response of the tower-base and blade-root bending moments is also well captured by the $\Delta x=10\,{\rm m}$ resolution inflow condition, especially during the eyewall hit (Fig. \ref{app:fig:loads_turb_freq}). Differences in the tower-base load variance contained near the natural frequency $f_{n,t}=0.17\,{\rm Hz}$ are $15~\%$, on average. For the blade-root moments, the load variance around $f_{n,t}=0.56\,{\rm Hz}$ is $25~\%$ and $10~\%$ larger for the $\Delta x = 5\,{\rm m}$ inflow for $t\approx1\,{\rm h}$ and $t\approx2\,{\rm h}$, respectively. Note that small differences in the frequency response are expected because both inflow conditions have statistically equivalent wind fields, not temporally equal flows. Nevertheless, the $\Delta x=10\,{\rm m}$ inflow excites the natural frequency of the tower and blade.

Our simulations indicate that the $\Delta x=10\,{\rm m}$ resolution LES is sufficient to capture the wind field structure and the associated turbine loads during the passage of the Category 4 hurricane, especially during the eyewall hit. The mean and turbulence profiles converge for the $\Delta x=10\,{\rm m}$ LES, compared to the $\Delta x=5\,{\rm m}$ simulation (Figs. \ref{app:fig:meanWinds} and \ref{app:fig:turb}). And, although the $\Delta x=5\,{\rm m}$ simulation captures more frequencies within the inertial subrange (Fig. \ref{app:fig:spctr}), the energy content in the flow at high frequencies decreases as $S_{ii}\sim f^{-5/3}$, making these contributions less significant. The turbine's response to the $\Delta x=10\,{\rm m}$ and $\Delta x=5\,{\rm m}$ LES inflow conditions is also similar. The load distribution of tower-base and blade-root bending moments generated by each inflow condition converge as the eyewall impacts the turbine position (Fig. \ref{app:fig:loads_quantiles}). In addition, the frequency response of the blade-root and tower-base bending moments is equivalent for both inflow conditions (Fig. \ref{app:fig:loads_turb_freq}). Notably, the $\Delta x=10\,{\rm m}$ inflow condition is sufficient to excite the first tower fore-aft and blade flapwise modes during the eyewall hit. Therefore, improvements in resolving smaller scales of turbulence provide diminishing returns for $\Delta x<10\,{\rm m}$ as those small eddies have low energy content and do not significantly change the loads on the turbine structure. However, it is possible that lower-intensity storms may require smaller grid spacing than $\Delta x=10\,{\rm m}$.

 \bibliographystyle{elsarticle-num} 
 \bibliography{references.bib}

@article{kapoor_hurricane_2020,
	title = {Hurricane eyewall winds and structural response of wind turbines},
	volume = {5},
	issn = {2366-7451},
	doi = {10.5194/wes-5-89-2020},
	pages = {89--104},
	number = {1},
	journal = {Wind Energy Science},
	shortjournal = {Wind Energ. Sci.},
	author = {Kapoor, Amber and Ouakka, Slimane and Arwade, Sanjay R. and Lundquist, Julie K. and Lackner, Matthew A. and Myers, Andrew T. and Worsnop, Rochelle P. and Bryan, George H.},
	urlyear = {2023-03-13},
	year = {2020},
	langid = {english},
}

@article{worsnop_gusts_2017,
	title = {Gusts and shear within hurricane eyewalls can exceed offshore wind turbine design standards},
	volume = {44},
	issn = {0094-8276, 1944-8007},
	doi = {10.1002/2017GL073537},
	pages = {6413--6420},
	number = {12},
	journal = {Geophysical Research Letters},
	shortjournal = {Geophys. Res. Lett.},
	author = {Worsnop, Rochelle P. and Lundquist, Julie K. and Bryan, George H. and Damiani, Rick and Musial, Walt},
	urlyear = {2023-03-13},
	year = {2017},
	langid = {english},
}

@article{sanGom_2023,
	title = {Wind {Fields} in {Category} 1–3 {Tropical} {Cyclones} {Are} {Not} {Fully} {Represented} in {Wind} {Turbine} {Design} {Standards}},
	volume = {128},
	issn = {2169-897X, 2169-8996},
	url = {https://agupubs.onlinelibrary.wiley.com/doi/10.1029/2023JD039233},
	doi = {10.1029/2023JD039233},
	language = {en},
	number = {16},
	urldate = {2024-05-08},
	journal = {JGR Atmospheres},
	author = {Sanchez Gomez, M. and Lundquist, J. K. and Deskos, G. and Arwade, S. R. and Myers, A. T. and Hajjar, J. F.},
	month = aug,
	year = {2023},
	pages = {e2023JD039233},
}

@techreport{iec-1_wind_2019,
	title = {Wind turbines - Part 1: Design requirements},
	number = {{IEC} 61400-1:2019},
	institution = {International Electrotechnical Commission},
	author = {{IEC}},
	year = {2019},
}

@techreport{iec-3_wind_2019,
	title = {Wind turbines – Part 3: Design requirements for offshore wind turbines},
	number = {{IEC} 61400-3-1:2019 },
	institution = {International Electrotechnical Commission},
	author = {{IEC}},
	year = {2019},
}

@article{zhang_asymmetric_2013,
	title = {Asymmetric {Hurricane} {Boundary} {Layer} {Structure} from {Dropsonde} {Composites} in {Relation} to the {Environmental} {Vertical} {Wind} {Shear}},
	volume = {141},
	issn = {0027-0644, 1520-0493},
	url = {http://journals.ametsoc.org/doi/10.1175/MWR-D-12-00335.1},
	doi = {10.1175/MWR-D-12-00335.1},
	language = {en},
	number = {11},
	urldate = {2024-05-10},
	journal = {Monthly Weather Review},
	author = {Zhang, Jun A. and Rogers, Robert F. and Reasor, Paul D. and Uhlhorn, Eric W. and Marks, Frank D.},
	month = nov,
	year = {2013},
	pages = {3968--3984},
}

@article{ren_largeeddy_2020,
	title = {Large‐Eddy Simulation of Idealized Hurricanes at Different Sea Surface Temperatures},
	volume = {12},
	issn = {1942-2466, 1942-2466},
	doi = {10.1029/2020MS002057},
	number = {9},
	journal = {Journal of Advances in Modeling Earth Systems},
	shortjournal = {J Adv Model Earth Syst},
	author = {Ren, Hehe and Dudhia, Jimy and Li, Hui},
	urlyear = {2023-03-13},
	year = {2020},
	langid = {english},
}

@article{ren_basic_2022,
	title = {The basic wind characteristics of idealized hurricanes of different intensity levels},
	volume = {225},
	issn = {01676105},
	doi = {10.1016/j.jweia.2022.104980},
	pages = {104980},
	journal = {Journal of Wind Engineering and Industrial Aerodynamics},
	shortjournal = {Journal of Wind Engineering and Industrial Aerodynamics},
	author = {Ren, Hehe and Dudhia, Jimy and Ke, Shitang and Li, Hui},
	urlyear = {2023-03-13},
	year = {2022},
	langid = {english},
}

@article{li_effects_2021,
	title = {Effects of {Roll} {Vortices} on the {Evolution} of {Hurricane} {Harvey} during {Landfall}},
	volume = {78},
	copyright = {http://www.ametsoc.org/PUBSReuseLicenses},
	issn = {0022-4928, 1520-0469},
	url = {https://journals.ametsoc.org/view/journals/atsc/78/6/JAS-D-20-0270.1.xml},
	doi = {10.1175/JAS-D-20-0270.1},
	number = {6},
	urldate = {2024-04-03},
	journal = {Journal of the Atmospheric Sciences},
	author = {Li, Xin and Pu, Zhaoxia and Gao, Zhiqiu},
	month = jun,
	year = {2021},
	pages = {1847--1867},
}

@article{zhang_characteristic_2011,
	title = {On the {Characteristic} {Height} {Scales} of the {Hurricane} {Boundary} {Layer}},
	volume = {139},
	issn = {0027-0644, 1520-0493},
	url = {https://journals.ametsoc.org/view/journals/mwre/139/8/mwr-d-10-05017.1.xml},
	doi = {10.1175/MWR-D-10-05017.1},
	number = {8},
	urldate = {2024-04-04},
	journal = {Monthly Weather Review},
	author = {Zhang, Jun A. and Rogers, Robert F. and Nolan, David S. and Marks, Frank D.},
	month = aug,
	year = {2011},
	pages = {2523--2535},
}

@techreport{skamarock_description_2019,
	title = {A Description of the Advanced Research {WRF} Model Version 4},
	institution = {{UCAR}/{NCAR}},
	author = {Skamarock, William C. and Klemp, Joseph B. and Dudhia, Jimy and Gill, David O. and Liu, Zhiquan and Berner, Judith and Wang, Wei and Powers, Jordan G. and Duda, Michael G. and Barker, Dale M. and Huang, Xiang-Yu},
	year = {2019},
	langid = {english},
	doi = {10.5065/1DFH-6P97},
}

@article{hong_wrf_2006,
	title = {The {WRF} Single-Moment 6-Class Microphysics Scheme ({WSM}6)},
	volume = {42},
	pages = {129--151},
	author = {Hong, S and Lim, J},
    journal = {Asia-pacific Journal of Atmospheric Sciences},
	year = {2006},
    volume={42},
    pages={129-151}
}

@article{skamarock_evaluating_2004,
	title = {Evaluating {Mesoscale} {NWP} {Models} {Using} {Kinetic} {Energy} {Spectra}},
	volume = {132},
	issn = {0027-0644, 1520-0493},
	url = {http://journals.ametsoc.org/doi/10.1175/MWR2830.1},
	doi = {10.1175/MWR2830.1},
	language = {en},
	number = {12},
	urldate = {2024-04-04},
	journal = {Mon. Wea. Rev.},
	author = {Skamarock, William C.},
	month = dec,
	year = {2004},
	pages = {3019--3032},
}

@techreport{gaertner2020iea,
  title={{IEA} wind {TCP} task 37: definition of the {IEA} 15-megawatt offshore reference wind turbine},
  author={Gaertner, Evan and Rinker, Jennifer and Sethuraman, Latha and Zahle, Frederik and Anderson, Benjamin and Barter, Garrett E and Abbas, Nikhar J and Meng, Fanzhong and Bortolotti, Pietro and Skrzypinski, Witold and others},
  year={2020},
  institution={National Renewable Energy Laboratory (NREL), Golden, CO (United States)}
}

@article{franklin_gps_2003,
	title = {{GPS} Dropwindsonde Wind Profiles in Hurricanes and Their Operational Implications},
	volume = {18},
	issn = {0882-8156, 1520-0434},
	doi = {10.1175/1520-0434(2003)018<0032:GDWPIH>2.0.CO;2},
	pages = {32--44},
	number = {1},
	journal = {Weather and Forecasting},
	shortjournal = {Wea. Forecasting},
	author = {Franklin, James L. and Black, Michael L. and Valde, Krystal},
	urldate = {2023-03-13},
	year = {2003},
	langid = {english},
}

@article{munoz-esparza_bridging_2014,
	title = {Bridging the Transition from Mesoscale to Microscale Turbulence in Numerical Weather Prediction Models},
	volume = {153},
	issn = {0006-8314, 1573-1472},
	url = {http://link.springer.com/10.1007/s10546-014-9956-9},
	doi = {10.1007/s10546-014-9956-9},
	pages = {409--440},
	number = {3},
	journal = {Boundary-Layer Meteorology},
	shortjournal = {Boundary-Layer Meteorol},
	author = {Muñoz-Esparza, Domingo and Kosović, Branko and Mirocha, Jeff and van Beeck, Jeroen},
	urldate = {2020-05-14},
	year = {2014},
	langid = {english}
}

@article{munoz-esparza_stochastic_2015,
	title = {A stochastic perturbation method to generate inflow turbulence in large-eddy simulation models: Application to neutrally stratified atmospheric boundary layers},
	volume = {27},
	issn = {1070-6631, 1089-7666},
	url = {http://aip.scitation.org/doi/10.1063/1.4913572},
	doi = {10.1063/1.4913572},
	shorttitle = {A stochastic perturbation method to generate inflow turbulence in large-eddy simulation models},
	pages = {035102},
	number = {3},
	journal = {Physics of Fluids},
	shortjournal = {Physics of Fluids},
	author = {Muñoz-Esparza, D. and Kosović, B. and van Beeck, J. and Mirocha, J.},
	urldate = {2020-05-14},
	year = {2015},
	langid = {english}
}

@article{jimenez_revised_2012,
	title = {A Revised Scheme for the {WRF} Surface Layer Formulation},
	volume = {140},
	issn = {0027-0644, 1520-0493},
	doi = {10.1175/MWR-D-11-00056.1},
	pages = {898--918},
	number = {3},
	journal = {Monthly Weather Review},
	author = {Jiménez, Pedro A. and Dudhia, Jimy and González-Rouco, J. Fidel and Navarro, Jorge and Montávez, Juan P. and García-Bustamante, Elena},
	urlyear = {2023-03-13},
	year = {2012},
	langid = {english},
}

@article{donelan_limiting_2004,
	title = {On the limiting aerodynamic roughness of the ocean in very strong winds},
	volume = {31},
	issn = {0094-8276},
	doi = {10.1029/2004GL019460},
	pages = {L18306},
	number = {18},
	journal = {Geophysical Research Letters},
	shortjournal = {Geophys. Res. Lett.},
	author = {Donelan, M. A.},
	urldate = {2023-03-13},
	year = {2004},
	langid = {english},
}

@article{willoughby_concentric_1982,
	title = {Concentric {Eye} {Walls}, {Secondary} {Wind} {Maxima}, and {The} {Evolution} of the {Hurricane} vortex},
	volume = {39},
	issn = {0022-4928, 1520-0469},
	url = {http://journals.ametsoc.org/doi/10.1175/1520-0469(1982)039<0395:CEWSWM>2.0.CO;2},
	doi = {10.1175/1520-0469(1982)039<0395:CEWSWM>2.0.CO;2},
	language = {en},
	number = {2},
	urldate = {2024-05-16},
	journal = {J. Atmos. Sci.},
	author = {Willoughby, H. E. and Clos, J. A. and Shoreibah, M. G.},
	month = feb,
	year = {1982},
	pages = {395--411},
}

@article{marks_structure_2008,
	title = {Structure of the {Eye} and {Eyewall} of {Hurricane} {Hugo} (1989)},
	volume = {136},
	issn = {1520-0493, 0027-0644},
	url = {http://journals.ametsoc.org/doi/10.1175/2007MWR2073.1},
	doi = {10.1175/2007MWR2073.1},
	language = {en},
	number = {4},
	urldate = {2024-05-16},
	journal = {Monthly Weather Review},
	author = {Marks, Frank D. and Black, Peter G. and Montgomery, Michael T. and Burpee, Robert W.},
	month = apr,
	year = {2008},
	pages = {1237--1259},
}

@article{lorsolo_estimation_2010,
	title = {Estimation and {Mapping} of {Hurricane} {Turbulent} {Energy} {Using} {Airborne} {Doppler} {Measurements}},
	volume = {138},
	issn = {1520-0493, 0027-0644},
	url = {http://journals.ametsoc.org/doi/10.1175/2010MWR3183.1},
	doi = {10.1175/2010MWR3183.1},
	language = {en},
	number = {9},
	urldate = {2024-05-16},
	journal = {Monthly Weather Review},
	author = {Lorsolo, Sylvie and Zhang, Jun A. and Marks, Frank and Gamache, John},
	month = sep,
	year = {2010},
	pages = {3656--3670},
}

@article{wang_long-term_2015,
	title = {A {Long}-{Term}, {High}-{Quality}, {High}-{Vertical}-{Resolution} {GPS} {Dropsonde} {Dataset} for {Hurricane} and {Other} {Studies}},
	volume = {96},
	issn = {0003-0007, 1520-0477},
	url = {https://journals.ametsoc.org/doi/10.1175/BAMS-D-13-00203.1},
	doi = {10.1175/BAMS-D-13-00203.1},
	language = {en},
	number = {6},
	urldate = {2024-05-16},
	journal = {Bulletin of the American Meteorological Society},
	author = {Wang, Junhong (June) and Young, Kate and Hock, Terry and Lauritsen, Dean and Behringer, Dalton and Black, Michael and Black, Peter G. and Franklin, James and Halverson, Jeff and Molinari, John and Nguyen, Leon and Reale, Tony and Smith, Jeff and Sun, Bomin and Wang, Qing and Zhang, Jun A.},
	month = jun,
	year = {2015},
	pages = {961--973},
}

@article{oguejiofor_role_2024,
	title = {The {Role} of {Turbulence} in an {Intense} {Tropical} {Cyclone}: {Momentum} {Diffusion}, {Eddy} {Viscosities}, and {Mixing} {Lengths}},
	issn = {0022-4928, 1520-0469},
	shorttitle = {The {Role} of {Turbulence} in an {Intense} {Tropical} {Cyclone}},
	url = {https://journals.ametsoc.org/view/journals/atsc/aop/JAS-D-23-0209.1/JAS-D-23-0209.1.xml},
	doi = {10.1175/JAS-D-23-0209.1},
	urldate = {2024-06-07},
	journal = {Journal of the Atmospheric Sciences},
	author = {Oguejiofor, Chibueze N. and Bryan, George H. and Rotunno, Richard and Sullivan, Peter P. and Richter, David H.},
	month = jun,
	year = {2024},
}

@techreport{asce_minimum_2021,
	address = {Reston},
	title = {Minimum {Design} {Loads} and {Associated} {Criteria} for {Buildings} and {Other} {Structures}},
	language = {eng},
	number = {ASCE/SEI 7-22},
	institution = {American Society of Civil Engineers},
	author = {ASCE},
	year = {2021},
	note = {OCLC: 1422497296},
}

@techreport{api_interim_2007,
	title = {Interim {Guidance} on {Hurricane} {Conditions} in the {Gulf} of {Mexico}},
	number = {API 2INT-MET},
	institution = {American Petroleum Institute},
	author = {API},
	year = {2007},
}

@techreport{api_recommended_2000,
	title = {Recommended {Practice} for {Planning}, {Designing} and {Constructing} {Fixed} {Offshore} {Platforms}—{Working} {Stress} {Design}},
	number = {API RP 2A-WSD},
	institution = {American Petroleum Institute},
	author = {API},
	year = {2000},
}

@article{mlawer_radiative_1997,
	title = {Radiative transfer for inhomogeneous atmospheres: {RRTM}, a validated correlated-k model for the longwave},
	volume = {102},
	issn = {01480227},
	url = {http://doi.wiley.com/10.1029/97JD00237},
	doi = {10.1029/97JD00237},
	shorttitle = {Radiative transfer for inhomogeneous atmospheres},
	pages = {16663--16682},
	issue = {D14},
	journal = {Journal of Geophysical Research: Atmospheres},
	shortjournal = {J. Geophys. Res.},
	author = {Mlawer, Eli J. and Taubman, Steven J. and Brown, Patrick D. and Iacono, Michael J. and Clough, Shepard A.},
	urldate = {2022-04-04},
	year = {1997},
	langid = {english}
}

@article{dudhia_numerical_1989,
	title = {Numerical Study of Convection Observed during the Winter Monsoon Experiment Using a Mesoscale Two-Dimensional Model},
	volume = {46},
	issn = {0022-4928, 1520-0469},
	url = {http://journals.ametsoc.org/doi/10.1175/1520-0469(1989)046<3077:NSOCOD>2.0.CO;2},
	doi = {10.1175/1520-0469(1989)046<3077:NSOCOD>2.0.CO;2},
	pages = {3077--3107},
	number = {20},
	journal = {Journal of the Atmospheric Sciences},
	shortjournal = {J. Atmos. Sci.},
	author = {Dudhia, Jimy},
	urldate = {2022-04-04},
	year = {1989},
	langid = {english}
}

@article{krayer_gust_1992,
	title = {Gust {Factors} {Applied} to {Hurricane} {Winds}},
	volume = {73},
	issn = {0003-0007, 1520-0477},
	url = {http://journals.ametsoc.org/doi/10.1175/1520-0477(1992)073<0613:GFATHW>2.0.CO;2},
	doi = {10.1175/1520-0477(1992)073<0613:GFATHW>2.0.CO;2},
	language = {en},
	number = {5},
	urldate = {2024-10-04},
	journal = {Bull. Amer. Meteor. Soc.},
	author = {Krayer, William R. and Marshall, Richard D.},
	month = may,
	year = {1992},
	pages = {613--618},
}

@article{durst_wind_1960,
	title = {Wind speeds over short periods of time},
	volume = {89},
	number = {1056},
	author = {Durst, C. S.},
	year = {1960},
	pages = {181--186},
}

@misc{cisl_rda_dsd084001,
 author = "{NOAA/NCEP}",
 title = "NCEP GFS 0.25 Degree Global Forecast Grids Historical Archive",
 publisher  = "Research Data Archive at the National Center for Atmospheric Research, Computational and Information Systems Laboratory",
 address = {Boulder CO},
 year  = "2015",
 url = "https://rda.ucar.edu/datasets/dsd084001/"
}

@inproceedings{stark_ostia_2007,
	address = {Aberdeen, UK},
	title = {{OSTIA} : {An} operational, high resolution, real time, global sea surface temperature analysis system},
	copyright = {https://doi.org/10.15223/policy-029},
	isbn = {978-1-4244-0634-0},
	shorttitle = {{OSTIA}},
	url = {https://ieeexplore.ieee.org/document/4302251/},
	doi = {10.1109/OCEANSE.2007.4302251},
	urldate = {2024-12-10},
	booktitle = {{OCEANS} 2007 - {Europe}},
	publisher = {IEEE},
	author = {Stark, John D. and Donlon, Craig J. and Martin, Matthew J. and McCulloch, Michael E.},
	month = jun,
	year = {2007},
	pages = {1--4},
}

@book{pope_turbulent_2015,
	address = {Cambridge},
	edition = {1. publ., 12. print},
	title = {Turbulent flows},
	isbn = {978-0-521-59886-6 978-0-521-59125-6},
	language = {eng},
	publisher = {Cambridge Univ. Press},
	author = {Pope, Stephen B.},
	year = {2015},
}

@article{stanislawski_effect_2023,
	title = {Effect of the integral length scales of turbulent inflows on wind turbine loads},
	volume = {217},
	issn = {09601481},
	url = {https://linkinghub.elsevier.com/retrieve/pii/S0960148123011333},
	doi = {10.1016/j.renene.2023.119218},
	language = {en},
	urldate = {2024-12-11},
	journal = {Renewable Energy},
	author = {Stanislawski, Brooke J. and Thedin, Regis and Sharma, Ashesh and Branlard, Emmanuel and Vijayakumar, Ganesh and Sprague, Michael A.},
	month = nov,
	year = {2023},
	pages = {119218},
}

@article{sanchezgomez_turbulenceresolving_2025,
	title = {Turbulence‐resolving simulations of {Hurricane} \textit{{Laura}} (2020): {Insights} into extreme winds and eyewall turbulence},
	issn = {0035-9009, 1477-870X},
	shorttitle = {Turbulence‐resolving simulations of {Hurricane} \textit{{Laura}} (2020)},
	url = {https://rmets.onlinelibrary.wiley.com/doi/10.1002/qj.70003},
	doi = {10.1002/qj.70003},
	language = {en},
	urldate = {2025-09-05},
	journal = {Quarterly Journal of the Royal Meteorological Society},
	author = {Sanchez‐Gomez, Miguel and Deskos, Georgios and Lundquist, Julie K.},
	month = aug,
	year = {2025},
	pages = {e70003},
}

@article{hallowell_hurricane_2018,
	title = {Hurricane risk assessment of offshore wind turbines},
	volume = {125},
	issn = {09601481},
	url = {https://linkinghub.elsevier.com/retrieve/pii/S0960148118302349},
	doi = {10.1016/j.renene.2018.02.090},
	language = {en},
	urldate = {2025-12-08},
	journal = {Renewable Energy},
	author = {Hallowell, Spencer T. and Myers, Andrew T. and Arwade, Sanjay R. and Pang, Weichiang and Rawal, Prashant and Hines, Eric M. and Hajjar, Jerome F. and Qiao, Chi and Valamanesh, Vahid and Wei, Kai and Carswell, Wystan and Fontana, Casey M.},
	month = sep,
	year = {2018},
	pages = {234--249},
}

@article{keim_spatiotemporal_2007,
	title = {Spatiotemporal {Patterns} and {Return} {Periods} of {Tropical} {Storm} and {Hurricane} {Strikes} from {Texas} to {Maine}},
	volume = {20},
	issn = {1520-0442, 0894-8755},
	url = {http://journals.ametsoc.org/doi/10.1175/JCLI4187.1},
	doi = {10.1175/JCLI4187.1},
	language = {en},
	number = {14},
	urldate = {2025-12-08},
	journal = {Journal of Climate},
	author = {Keim, Barry D. and Muller, Robert A. and Stone, Gregory W.},
	month = jul,
	year = {2007},
	pages = {3498--3509},
}

@article{pandey_typhoon_2022,
	title = {Typhoon strength rising in the past four decades},
	volume = {36},
	issn = {22120947},
	url = {https://linkinghub.elsevier.com/retrieve/pii/S2212094722000329},
	doi = {10.1016/j.wace.2022.100446},
	language = {en},
	urldate = {2025-12-08},
	journal = {Weather and Climate Extremes},
	author = {Pandey, Ravi Shankar and Liou, Yuei-An},
	month = jun,
	year = {2022},
	pages = {100446},
}

@article{chen_assessing_2018,
	title = {Assessing typhoon damages to {Taiwan} in the recent decade: return period analysis and loss prediction},
	volume = {91},
	issn = {0921-030X, 1573-0840},
	shorttitle = {Assessing typhoon damages to {Taiwan} in the recent decade},
	url = {http://link.springer.com/10.1007/s11069-017-3159-x},
	doi = {10.1007/s11069-017-3159-x},
	language = {en},
	number = {2},
	urldate = {2025-12-08},
	journal = {Natural Hazards},
	author = {Chen, Chia-Jeng and Lee, Tsung-Yu and Chang, Che-Min and Lee, Jun-Yi},
	month = mar,
	year = {2018},
	pages = {759--783},
}

@article{huang_long-term_2019,
	title = {Long-term and inter-annual variations of tropical cyclones affecting {Taiwan} region},
	volume = {30},
	issn = {23524855},
	url = {https://linkinghub.elsevier.com/retrieve/pii/S2352485519302373},
	doi = {10.1016/j.rsma.2019.100721},
	language = {en},
	urldate = {2025-12-08},
	journal = {Regional Studies in Marine Science},
	author = {Huang, Weinan and Dong, Sheng},
	month = jul,
	year = {2019},
	pages = {100721},
}

@article{fang_estimation_2020,
	title = {Estimation of tropical cyclone wind hazards in coastal regions of {China}},
	volume = {20},
	copyright = {https://creativecommons.org/licenses/by/4.0/},
	issn = {1684-9981},
	url = {https://nhess.copernicus.org/articles/20/1617/2020/},
	doi = {10.5194/nhess-20-1617-2020},
	language = {en},
	number = {6},
	urldate = {2025-12-08},
	journal = {Natural Hazards and Earth System Sciences},
	author = {Fang, Genshen and Zhao, Lin and Cao, Shuyang and Zhu, Ledong and Ge, Yaojun},
	month = jun,
	year = {2020},
	pages = {1617--1637},
}

@inproceedings{ishihara2005analysis,
  title={An analysis of damaged wind turbines by typhoon Maemi in 2003},
  author={Ishihara, Takeshi and Yamaguchi, Atsushi and Takahara, Keiji and Mekaru, Takehiro and Matsuura, Shinich},
  booktitle={Proceedings of the sixth asia-pacific conference on wind engineering},
  pages={1413--1428},
  year={2005}
}

@article{takahara2004damages,
  title={Damages of wind turbine on Miyakojima Island by Typhoon Maemi in 2003},
  author={Takahara, Keiji and MEKARU, Takehiro and SHINJO, Fumihiro and ISHIHARA, Takeshi and YAMAGUCHI, Atsushi and MATSUURA, Shinich},
  booktitle={European Wind Energy Conference \& Exhibition 2004},
  year={2004}
}

@article{kwasinski_hurricane_2019,
	title = {Hurricane {Maria} {Effects} on {Puerto} {Rico} {Electric} {Power} {Infrastructure}},
	volume = {6},
	copyright = {https://ieeexplore.ieee.org/Xplorehelp/downloads/license-information/OAPA.html},
	issn = {2332-7707},
	url = {https://ieeexplore.ieee.org/document/8644031/},
	doi = {10.1109/JPETS.2019.2900293},
	number = {1},
	urldate = {2025-12-08},
	journal = {IEEE Power and Energy Technology Systems Journal},
	author = {Kwasinski, Alexis and Andrade, Fabio and Castro-Sitiriche, Marcel J. and O'Neill-Carrillo, Efrain},
	month = mar,
	year = {2019},
	pages = {85--94},
}

@inproceedings{kwasinski_effects_2018,
	address = {Paris},
	title = {Effects of {Hurricane} {Maria} on {Renewable} {Energy} {Systems} in {Puerto} {Rico}},
	isbn = {978-1-5386-5982-3},
	url = {https://ieeexplore.ieee.org/document/8566922/},
	doi = {10.1109/ICRERA.2018.8566922},
	urldate = {2025-12-08},
	booktitle = {2018 7th {International} {Conference} on {Renewable} {Energy} {Research} and {Applications} ({ICRERA})},
	publisher = {IEEE},
	author = {Kwasinski, Alexis},
	month = oct,
	year = {2018},
	pages = {383--390},
}

@article{sroka_organized_2021,
	title = {Organized kinetic energy backscatter in the hurricane boundary layer from radar measurements},
	volume = {924},
	issn = {0022-1120, 1469-7645},
	url = {https://www.cambridge.org/core/product/identifier/S0022112021006327/type/journal_article},
	doi = {10.1017/jfm.2021.632},
	language = {en},
	urldate = {2024-11-08},
	journal = {Journal of Fluid Mechanics},
	author = {Sroka, Sydney and Guimond, S.R.},
	month = oct,
	year = {2021},
	pages = {A21},
}

@misc{muller_can_2025,
	title = {Can the {Mann} model describe the typhoon turbulence?},
	copyright = {https://creativecommons.org/licenses/by/4.0/},
	url = {https://wes.copernicus.org/preprints/wes-2025-7/},
	doi = {10.5194/wes-2025-7},
	urldate = {2025-12-08},
	author = {Müller, Sara and Guo Larsén, Xiaoli and Hu, Fei},
	month = feb,
	year = {2025},
}

@article{mann_spatial_1994,
	title = {The spatial structure of neutral atmospheric surface-layer turbulence},
	volume = {273},
	copyright = {https://www.cambridge.org/core/terms},
	issn = {0022-1120, 1469-7645},
	url = {https://www.cambridge.org/core/product/identifier/S0022112094001886/type/journal_article},
	doi = {10.1017/S0022112094001886},
	language = {en},
	urldate = {2025-12-08},
	journal = {Journal of Fluid Mechanics},
	author = {Mann, Jakob},
	month = aug,
	year = {1994},
	pages = {141--168},
}

@article{sanchez_gomez_toward_2025,
	title = {Toward {Understanding} the {Differences} between {Mesoscale} and {Large}-{Eddy} {Simulations} of {Tropical} {Cyclones}},
	volume = {82},
	copyright = {https://creativecommons.org/licenses/by/4.0/},
	issn = {0022-4928, 1520-0469},
	url = {https://journals.ametsoc.org/view/journals/atsc/82/7/JAS-D-24-0131.1.xml},
	doi = {10.1175/JAS-D-24-0131.1},
	number = {7},
	urldate = {2025-09-05},
	journal = {Journal of the Atmospheric Sciences},
	author = {Sanchez Gomez, Miguel and Deskos, Georgios and Lundquist, Julie K.},
	month = jul,
	year = {2025},
	pages = {1293--1315},
}

@article{nolan_limits_2014,
	title = {On the {Limits} of {Estimating} the {Maximum} {Wind} {Speeds} in {Hurricanes}},
	volume = {142},
	issn = {0027-0644, 1520-0493},
	url = {http://journals.ametsoc.org/doi/10.1175/MWR-D-13-00337.1},
	doi = {10.1175/MWR-D-13-00337.1},
	language = {en},
	number = {8},
	urldate = {2025-12-09},
	journal = {Monthly Weather Review},
	author = {Nolan, David S. and Zhang, Jun A. and Uhlhorn, Eric W.},
	month = aug,
	year = {2014},
	pages = {2814--2837},
}

@article{bryan_simple_2017,
	title = {A {Simple} {Method} for {Simulating} {Wind} {Profiles} in the {Boundary} {Layer} of {Tropical} {Cyclones}},
	volume = {162},
	issn = {0006-8314, 1573-1472},
	url = {http://link.springer.com/10.1007/s10546-016-0207-0},
	doi = {10.1007/s10546-016-0207-0},
	language = {en},
	number = {3},
	urldate = {2024-10-09},
	journal = {Boundary-Layer Meteorology},
	author = {Bryan, George H. and Worsnop, Rochelle P. and Lundquist, Julie K. and Zhang, Jun A.},
	month = mar,
	year = {2017},
	pages = {475--502},
}

@article{keith_learning_2021,
	title = {Learning the structure of wind: {A} data-driven nonlocal turbulence model for the atmospheric boundary layer},
	volume = {33},
	issn = {1070-6631, 1089-7666},
	shorttitle = {Learning the structure of wind},
	url = {https://pubs.aip.org/pof/article/33/9/095110/1064702/Learning-the-structure-of-wind-A-data-driven},
	doi = {10.1063/5.0064394},
	language = {en},
	number = {9},
	urldate = {2025-12-09},
	journal = {Physics of Fluids},
	author = {Keith, B. and Khristenko, U. and Wohlmuth, B.},
	month = sep,
	year = {2021},
	pages = {095110},
}

@article{munozesparza_fasteddy_2022,
	title = {The {FastEddy} $^{\textrm{®}}$ {Resident}‐{GPU} {Accelerated} {Large}‐{Eddy} {Simulation} {Framework}: {Moist} {Dynamics} {Extension}, {Validation} and {Sensitivities} of {Modeling} {Non}‐{Precipitating} {Shallow} {Cumulus} {Clouds}},
	volume = {14},
	issn = {1942-2466, 1942-2466},
	shorttitle = {The {FastEddy} $^{\textrm{®}}$ {Resident}‐{GPU} {Accelerated} {Large}‐{Eddy} {Simulation} {Framework}},
	doi = {10.1029/2021MS002904},
        language = {en},
	number = {4},
	journal = {J Adv Model Earth Syst},
	author = {Muñoz‐Esparza, Domingo and Sauer, Jeremy A. and Jensen, Anders A. and Xue, Lulin and Grabowski, Wojciech W.},
	year = {2022},
	pages = {e2021MS002904},
}

@article{sauer_fasteddy_2020,
	title = {The {FastEddy}® {Resident}‐{GPU} {Accelerated} {Large}‐{Eddy} {Simulation} {Framework}: {Model} {Formulation}, {Dynamical}‐{Core} {Validation} and {Performance} {Benchmarks}},
	volume = {12},
	issn = {1942-2466, 1942-2466},
	shorttitle = {The {FastEddy}® {Resident}‐{GPU} {Accelerated} {Large}‐{Eddy} {Simulation} {Framework}},
	doi = {10.1029/2020MS002100},
        language = {en},
	number = {11},
	urldate = {2024-02-13},
	journal = {J Adv Model Earth Syst},
	author = {Sauer, Jeremy A. and Muñoz‐Esparza, Domingo},
	year = {2020},
	pages = {e2020MS002100},
}

@book{garratt_atmospheric_1999,
	address = {Cambridge},
	edition = {1. paperback ed. (with corr.)},
	series = {Cambridge atmospheric and space science series},
	title = {The atmospheric boundary layer},
	isbn = {978-0-521-46745-2 978-0-521-38052-2},
	language = {eng},
	publisher = {Cambridge University Press},
	author = {Garratt, John R.},
	year = {1999},
}

@article{anderson_development_2022,
	title = {Development of an open-source segmented blade design tool},
	volume = {2265},
	issn = {1742-6588, 1742-6596},
	url = {https://iopscience.iop.org/article/10.1088/1742-6596/2265/3/032023},
	doi = {10.1088/1742-6596/2265/3/032023},
	number = {3},
	urldate = {2025-12-09},
	journal = {Journal of Physics: Conference Series},
	author = {Anderson, Benjamin and Bortolotti, Pietro and Johnson, Nick},
	month = may,
	year = {2022},
	pages = {032023},
}

@techreport{griffith_structural_2009,
	address = {Albuquerque, New Mexico},
	title = {Structural {Dynamics} {Analysis} and {Model} {Validation} of {Wind} {Turbine} {Structures}},
	url = {https://www.osti.gov/servlets/purl/1142154},
	number = {SAND2009-2192C},
	institution = {Sandia National Laboratories},
	author = {Griffith, D. T.},
	year = {2009},
}

@techreport{jonkman_fast_2005,
	title = {{FAST} {User}’s {Guide}},
	url = {https://docs.nlr.gov/docs/fy06osti/38230.pdf},
	number = {NREL/TP-500-38230},
	author = {Jonkman, Jason and Buhl Jr, M.L.},
	year = {2005},
}

@article{kaimal_1972,
author = {Kaimal, J. C. and Wyngaard, J. C. and Izumi, Y. and Coté, O. R.},
title = {Spectral characteristics of surface-layer turbulence},
journal = {Quarterly Journal of the Royal Meteorological Society},
volume = {98},
number = {417},
pages = {563-589},
doi = {https://doi.org/10.1002/qj.49709841707},
url = {https://rmets.onlinelibrary.wiley.com/doi/abs/10.1002/qj.49709841707},
eprint = {https://rmets.onlinelibrary.wiley.com/doi/pdf/10.1002/qj.49709841707},
year = {1972}
}







\end{document}